\documentclass[useAMS,usenatbib]{mn2e} 
\usepackage[english]{babel}
\usepackage{journals}
\usepackage[pdftex]{graphicx,color} 
\usepackage[latin1]{inputenc}
\usepackage{graphics} 
\usepackage{amsfonts} 
\usepackage{amsmath}
\usepackage{multicol} 
\usepackage{layout} 
\usepackage{amssymb}
\usepackage[colorlinks=true,pdfstartview=FitV,linkcolor=red,citecolor=blue,urlcolor=magenta]{hyperref}

\title[Halo shapes from MultiDark \textit{Planck} simulations]{On the shape of dark matter haloes from MultiDark \textit{Planck} simulations}
\author[Vega-Ferrero, J. et al. 2017]
{\parbox{\textwidth}{
	Jes\'us Vega-Ferrero$^{1,2,3}$\thanks{E-mail: \href{mailto:astrovega@gmail.com} {astrovega@gmail.com}},
	Gustavo Yepes$^{2,4}$ and  Stefan Gottl\"ober$^5$}\\ \\ 
$^1$ Department of Physics and Astronomy, University of Pennsylvania, 209 S. 33rd St, Philadelphia, PA 19104, USA\\
$^2$ Departamento de F\'{\i}sica Te\'orica, Universidad Aut\'onoma de Madrid,  E-28049 Madrid, Spain\\
$^3$ LERMA, CNRS UMR 8112, Observatoire de Paris, 61 Avenue de l'Observatoire, F-75014 Paris, France\\
$^4$  ASTRO UAM, UAM Unidad Asociada CSIC\\
$^5$ Leibniz-Institut f\"ur Astrophysik, D-14482 Potsdam, Germany
}

\def\aap{A\&A}
\def\apjl{ApJ Letters}
\def\apjs{ApJS}

\AtBeginShipout{%
  \ifnum\value{page}>1 %
    \typeout{* Additional boxing of page `\thepage'}%
    \setbox\AtBeginShipoutBox=\hbox{\copy\AtBeginShipoutBox}%
  \fi
}

\begin{document}
\date{}
\maketitle
\label{firstpage}
\pagerange{\pageref{firstpage}--\pageref{lastpage}} \pubyear{2017}

\begin{abstract}
The halo  shape  plays a central role in determining important observational properties of the haloes such as mass, concentration and lensing cross-sections. The triaxiality of lensing galaxy clusters has a substantial impact on the distribution of the largest Einstein radii, while weak lensing techniques are sensitive to the intrinsic halo ellipticity. In this work, we provide scaling relations for the shapes of dark matter haloes as a function of mass (peak height) and redshift over more than four orders of magnitude in halo masses, namely from $10^{11.5}$ to $10^{15.8}~h^{-1}~$M$_\odot$. We have analysed four dark matter only simulations from the MultiDark cosmological simulation suite with  more than 56 billion particles within boxes of 4.0, 2.5, 1.0 and 0.4 $h^{-1}$Gpc size assuming \textit{Planck} cosmology. The dark matter haloes have been identified in the simulations  using the {\sc rockstar} halo finder, which also determines the axis ratios in terms of the diagonalization of the inertia tensor. In order to infer the shape for a hypothetical halo of a given mass at a given redshift, we provide fitting functions to the minor-to-major and intermediate-to-major axis ratios as a function of the peak height.

\end{abstract}


\section{Introduction}
\label{sec:intro}

According to  the current paradigm of structure formation isolated galaxies, groups and clusters of galaxies are hosted by dark matter haloes of increasing mass. Quantum fluctuations generated during an early, inflationary epoch of the Universe become  classical density fluctuations: the seed of today's  observed large scale structures. In an expanding Universe dominated by weakly interacting dark matter, these fluctuations grow with time due to the gravitational  instability that leads to the formation of  self-gravitating dark matter haloes.  The  first haloes emerge from  small scale fluctuations and progressively merge to form later generations of more massive haloes, where  groups and clusters of galaxies are hosted. These  haloes can   be associated with the peaks in the primordial Gaussian density field of dark matter, and their mass distributions can be derived in terms of the statistics of Gaussian random fields and the spherical collapse model of halo formation \citep{Press-Schechter1974}.

Contrary to the model assumption the mass accretion on to haloes comes along a preferential direction (mostly along a filament) and tends to be clumpy, i.e. massive haloes form by merging with smaller haloes.  For these reasons, one should not expect that dark matter haloes show a spherical shape if their relaxation times are longer than the time between mergers or their accretion events occur from a preferred  direction. For the most massive haloes that host clusters of galaxies, observational studies based on optical and density maps \citep{carter80,bingelli82}, X-ray data \citep{fabricant84,buote92,buote96,kawahara,lau2013}, Sunyaev Zel'dovich pressure maps \citep{sayers2011a}, strong and weak gravitational lensing methods \citep{soucail87,evans09,oguri10,oguri12}, indicate that  cluster-size dark matter haloes are not spherical at all.

Additionally, the study of dark matter haloes formed in cosmological simulations has shown that their shapes are well described by a triaxial model \citep{frenk88,dubinski91,warren92,cole96,JS02,Bailin2005,hopkins05,Kasun2005,Allgood2006,paz06,Bett2007,munozcuartas,gao12,Schneider2012,Despali2013}. Clusters of galaxies seem to be less spherical towards their centres \citep{warren92, JS02,Bailin2005, Allgood2006,  Schneider2012}, probably  due  to interactions  with other clusters or groups of galaxies. Besides, triaxiality of clusters increases with increasing both the mass and the redshift of the cluster \citep{JS02,Allgood2006,munozcuartas}, essentially because they are affected by the direction of the last major merger or the presence of filaments around them, and because massive haloes were formed later on \citep{Despali2014}. This scenario has been confirmed by the recent work of  \citet{Bonamigo2015} on the shape of simulated dark matter haloes extracted from the Millennium XXL simulation \citep{Angulo2012}.

It is well known that the triaxiality of  dark matter haloes is particularly important for the study of clusters of galaxies. In fact, the triaxiality of dark matter haloes is supposed to play a central role in several important properties: cluster mass and concentration, inner slope of the dark matter density profile and strong lensing cross-sections \citep{1998Bartelmann,2000Meneghetti,2001Meneghetti,2005Oguri,Giocoli2012a,Giocoli2012b,Wojtak2013}. In particular, the triaxiality of lensing galaxy clusters has a substantial impact on the distribution of the largest Einstein radii (\citealt{Waizmann12,Waizmann14}, Vega-Ferrero et al., in preparation). Moreover, semi-analytic models designed to derive the gravitational lensing properties in galaxy clusters generally described the mass distribution by a triaxial model as proposed by \citet{JS02}. The results presented by \citet{JS02} were  obtained from numerical simulations that hardly contained any halo above $10^{14}h^{-1}$M$_\odot$. Therefore, arc statistics from galaxy clusters inferred by these models \citep[see][for some examples]{Giocoli2012a,Redlich12} rely on extrapolations from lower mass haloes. In order to study the impact of the triaxiality, \citet{Waizmann12} introduced a cut-off in the distribution of axis ratios of \citet{JS02} to remove extreme axis ratios, finding that the impact of the tail of the axis ratio distribution on the largest Einstein radii distribution is substantial. However, due to the limited knowledge of the statistics of haloes with extremely small axis ratios, it is not possible to clearly define a proper choice of the cut-off (if present) until the triaxiality distributions of large halo samples (up to cluster masses) are derived  from large volume  numerical simulations.

Apart from some analyses over few  massive clusters \citep{munozcuartas,Schneider2012,Despali2014,Groener2014},  the work of \citet{Bonamigo2015} is the only one that presents statistically significant predictions on the shapes of  haloes with masses   above $3\times 10^{14}~h^{-1}$M$_\odot$. The MultiDark suite of cosmological simulations \citep{Klypin2016} -- in particular the Huge-MultiDark simulation ($\textsc{hmd}$),  containing almost 70 billion particles in a 4$h^{-1}$Gpc box volume -- allows us to study in detail the shape of simulated dark matter haloes in the cluster mass range with unprecedented statistics and in the most up-to-date cosmological parameters  consistent with the recent results from the \textit{Planck} Collaboration. 

Moreover, weak gravitational lensing \citep[see][for more details]{Bartelmann2001} has been successfully used in a wide variety of applications: from the mass calibration of galaxy clusters \citep{2015MNRAS.447.1304F,2015MNRAS.449..685H,2015MNRAS.451.1460K,2016MNRAS.457.1522A,2016ApJ...817...24M,2016ApJ...818L..25M} to the measurement of the ellipticity of dark matter haloes \citep{2000ApJ...538L.113N,evans09,oguri10,oguri12,2015MNRAS.454.1432S}. Since weak lensing is sensitive to halo ellipticity (or triaxiality), a detailed study of the intrinsic shapes of galaxy-scale simulated haloes is essential for present and future weak lensing studies, especially in the time of ``precision cosmology".

In this paper, we aim at characterizing  the shape of simulated dark matter haloes over more than four orders of magnitude in halo masses.  To this end:
\begin{enumerate}
\item we have analysed four high resolution dark matter only simulations within boxes of 4.0, 2.5, 1.0 and 0.4 $~h^{-1}$Gpc
\item we examine the shapes of haloes with more than 3,000 particles and provide predictions within the halo mass range: $3 \times 10^{11} \lesssim$ M $\lesssim 8 \times 10^{15}~h^{-1}$M$_\odot$;
\item we investigate the radial dependence of the halo shapes by measuring the axis ratios: at  virial radius (defined at an overdensity $\Delta_{\rm{vir}} \rho_{b}$) and at an overdensity of $500 \rho_{c}$, along with the alignment of the major axis at these overdensities;
\item additionally, we examine the probability distribution of the minima (minor-to-major) axis ratio.
\end{enumerate}

This paper is organised as follows. In Section~\ref{sec:catalogs}, we describe the set of simulations. In Section~\ref{sec:shapes}, we describe the different methods used to measure the shape of dark matter haloes and present our results for the four simulated boxes. In Section~\ref{sec:comparison} we compare our findings for the minor-to-major axis ratios as a function of halo mass with previous studies. Finally, in Section~\ref{sec:conclusions} we summarize our results and conclusions.

\section{Halo catalogues}
\label{sec:catalogs}

We derived the shape of simulated dark matter haloes over more than four orders of magnitude in halo masses. To this end we analysed four dark matter only simulations from the MultiDark cosmological simulation suite. This set of simulations consist on four cosmological volumes of 4.0, 2.5, 1.0 and 0.4 $~h^{-1}$Gpc cube sides, which are labelled here as Huge-MultiDark ($\textsc{hmd}$), Big-MultiDark ($\textsc{bmd}$), New-MultiDark ($\textsc{nmd}$) and Small-MultiDark ($\textsc{smd}$), respectively. The largest simulation volume (i.e. the $\textsc{hmd}$) contains 4096$^3$ particles, while the other  simulations are performed with 3840$^3$ particles. All the simulations were carried out with {\sc l-gadget-2} code \citep{Springel2005} and were performed assuming the same $\Lambda$ cold dark matter ($\Lambda$CDM) cosmological model with \textit{Planck} cosmological parameters \citep{planck_params}: ($\Omega_{\rm{M}}$, $\Omega_{\rm{b}}$, $\Omega_{\rm{\Lambda}}$, $\rm{\sigma_8}$, $n_s$, $h$) = (0.307, 0.048, 0.693, 0.829, 0.96, 0.678). For these reasons, the MultiDark suite constitutes a coherent and homogeneous data set of cosmological simulations. The particle masses along with the parameters for each simulation are shown in Table \ref{tb:sims1}. All the information concerning these simulations is available at the MultiDark data base\footnote{\url{http://www.multidark.org}}
 and the CosmoSim data base\footnote{\url{https://www.cosmosim.org}}.

\begin{table*}
\begin{center}
\caption{Numerical parameters for the simulations. The columns correspond to: the simulation identifier; the size of the simulated box in $h^{-1}$Gpc; the number of particles and the mass per simulation particle $m_{\rm{p}}$ in units $h^{-1}$M$_\odot$; the Plummer equivalent gravitational softening length $\epsilon$ in units of physical $h^{-1}$kpc; the number of haloes at $z=0$; the number of relaxed haloes at $z=0$ with more than 3,000 particles; the fraction of relaxed haloes at $z=0$.}
\label{tb:sims1}
\begin{tabular}{cccccccc}
\hline
Simulation & Box & Particles & m$_p$ & $\epsilon$ & $N_{\rm{h}}$ ($z=0$) & $N_{\rm{rel}}$ ($z=0$) & $f_{\rm{rel}}$  ($z=0$)\\
\hline
$\textsc{smd}$ & 0.4 & 3840$^3$ & 9.6$\times$10$^7$ & 1.5 & 814977 & 638414 & 0.78\\
$\textsc{nmd}$ & 1.0 & 3840$^3$ & 1.5$\times$10$^9$ & 5 & 1004645 & 720799 & 0.72\\
$\textsc{bmd}$ & 2.5 & 3840$^3$ & 2.4$\times$10$^{10}$ & 10 & 692647 & 432815 & 0.62\\
$\textsc{hmd}$ & 4.0 & 4096$^3$ & 7.9$\times$10$^{10}$ & 25 & 387849 & 219567 & 0.57\\
\end{tabular}
\end{center}
\end{table*}

Dark matter haloes are identified using the {\sc rockstar} halo finder \citep{Behroozi2013}. From the {\sc rockstar} halo catalogues we select all the distinct haloes at four different redshifts, $z = (0.00, 0.33, 0.66, 1.00)$. Among other parameters, the {\sc rockstar} halo finder provides virial masses and the length of the minor, intermediate and major axis. Thereafter we use the definition of the virial mass as the mass inside the radius encompassing a given density contrast $\Delta_{\rm{vir}} \rho_b$ and $500 \rho_c$, where $\rho_{b}$ and $\rho_{c}$ are the mean and the critical density of the Universe, respectively. The overdensity $\Delta_{\rm vir}$ is computed using the approximation given by \citet{BN1998}. For the given cosmology, $\Delta_{\rm vir} \approx 333$ times the mean density of the Universe at $z=0$.

\begin{figure}
\includegraphics[width=8.5cm,angle=0.0]{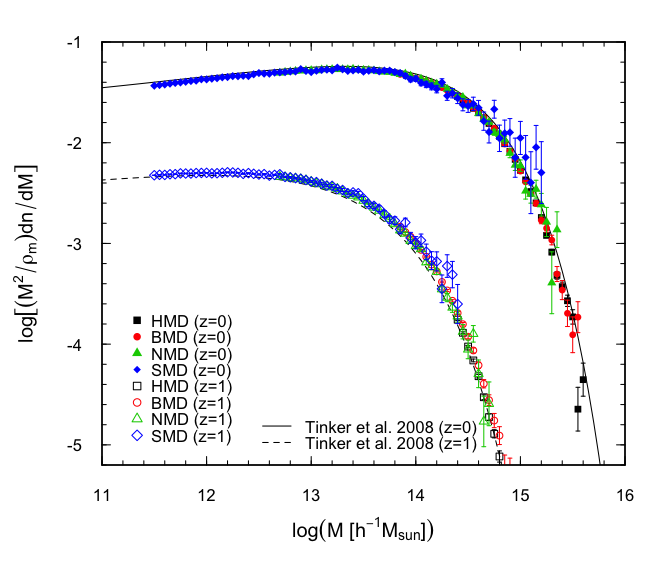}
\caption{Halo mass function in simulations with the \textit{Planck} cosmological parameters. Different points correspond to the different simulations at $z=0$ and 1. Error bars are computed considering that the error in the number of haloes is $\Delta n = \sqrt{n}$. The full curves correspond to the \citet{Tinker2008} mass function. It gives an excellent fit to $z = 0$ results and slightly underestimates the mass function at $z=1$.}
\label{fig:mass_functions}
\end{figure}

To ensure a good resolution we limited our analysis to distinct haloes with more than 3,000 particles within $\rm{R_{vir}}$. A sample build in this way also constitutes a complete volume limited sample of the $\textsc{hmd}$, $\textsc{bmd}$, $\textsc{nmd}$ and $\textsc{smd}$ box volumes. In Fig.~\ref{fig:mass_functions} we show the halo mass function for the different simulations with the \textit{Planck} cosmology. The \citet{Tinker2008} mass function provides an excellent fit to $z=0$ results and slightly underestimates the mass function at $z=1$. 

\begin{figure}
\includegraphics[width=8.5cm,angle=0.0]{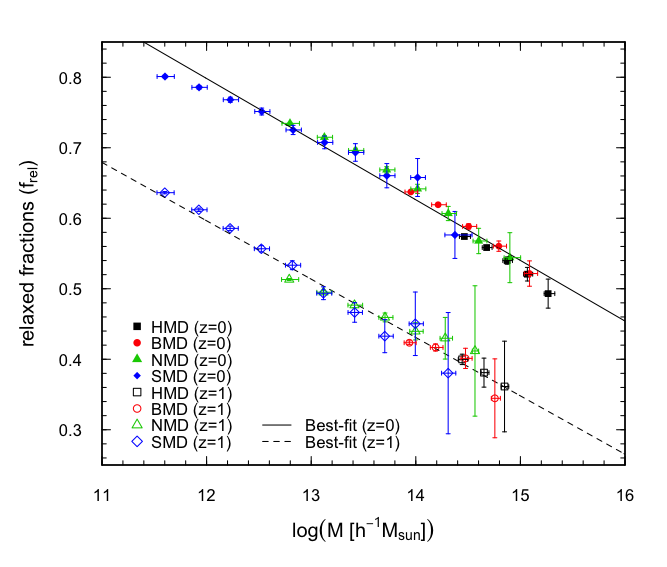}
\caption{Fraction of relaxed haloes as a function of the halo mass. Different points correspond to the median relaxed fraction for the different simulated boxes at $z=0$ and 1. Vertical error bars are computed considering that the error in the number of haloes is $\Delta n = \sqrt{n}$, while horizontal error bars enclosed the Q1 and Q3 quartiles. The solid and dashed lines indicate the linear best fit to the data points for $z=0$ and 1, respectively.}
\label{fig:relaxed_fractions}
\end{figure}

Additionally, the modelling of non spherical objects is not simple and can lead to uncertain results. Since highly elongated haloes are mostly due to unrelaxed or merging haloes, we divide our sample in relaxed and unrelaxed haloes. Relaxed haloes are defined by these three criteria \citep{Klypin2016}:
\begin{equation}
r_{\rm{off}} < 0.07 \;, \lambda < 0.07 \;, 2K/ |W| < 1.5\;,
\label{eq:rlx_conds}
\end{equation}
where $\lambda$ is the spin parameter as introduced by \citet{Peebles1969}, $2K/ |W|$ is the virial parameter and $r_{\rm{off}} = | r_{\rm{cm}} - \textbf{r}_{\phi} | / \rm{R_{vir}}$ is defined as the offset between the centre of mass ($r_{\rm{cm}}$) and the centre of the potential ($\textbf{r}_{\phi}$), in units of the virial radius. The fraction of relaxed haloes as a function of the halo mass for the different simulations are shown in Fig.~\ref{fig:relaxed_fractions}. At redshift $z=0$, this fraction drops from 80 per cent down to 50 per cent in the mass range considered in this work. The merger rate of massive haloes is usually larger and more violent compared to lower mass haloes, which explains the negative slope in the relaxed fraction. At $z=1$, this fraction is smaller in about 25 per cent than that at $z=0$ within the whole mass range. The fraction of relaxed haloes at $z=0$ for each simulation are shown in Table \ref{tb:sims1}.

\section{Halo shapes}
\label{sec:shapes}

The common procedure to determine the shape of haloes is to model them as ellipsoids. The length of the minor, intermediate and major axis ($c \leqslant b \leqslant a$) of the ellipsoid can be computed by diagonalizing the inertia tensor. In this work we determine the axis ratios using two forms of the inertia tensor that are commonly used in the literature: the unweighted inertia tensor,

\begin{equation} 
I_{ij} \equiv \sum_{n} x_{i,n} x_{j,n}\;;
\label{eq:inertia}  
\end{equation}
and the weighted (or reduced) inertia tensor,

\begin{equation} 
\tilde{I}_{ij} \equiv \sum_{n} \frac{x_{i,n} x_{j,n}}{r_n^2}\;,
\label{eq:inertia_w}  
\end{equation}
where
\begin{equation} r_n = \sqrt{x_n^2 + y_n^2/q^2 + z_n^2/s^2}
\end{equation}
is the elliptical distance in the eigenvector coordinate system from the halo centre to the $n$th particle. The square root of the eigenvalues obtained from the diagonalization of both expressions provides the axis ratios ($c \leqslant b \leqslant a$), while the corresponding eigenvectors determine their orientations. Halo centres are identified as the densest point in phase space, which normally corresponds to the densest point in position space.

The minor-to-major and intermediate-to-major axis ratios are defined as $s \equiv c/a$ and $q \equiv b/a$, respectively. Hereafter, we denote the virial axis ratios measured at an overdensity $\Delta_{\rm{vir}} \rho_b$ (i.e. $\rm{R_{vir}}$) as $s$ and $q$. Additionally, the axis ratios measured at an overdensity of $500 \rho_c$ ($\rm{R_{500}}$) are denoted as $\textit{s}_{500}$ and $\textit{q}_{500}$.

The procedure used to recover the shape begins by determining the inertia tensor with $s = 1$ and $q=1$, including all the particles inside a sphere of a given overdensity. Subsequently, new values for $s$ and $q$ are computed as the volume analysis is deformed along the eigenvectors (proportionally to the eigenvalues), while keeping the major axis equal to the original spherical radius. After the deformations of the original spherical region, the inertia tensor is calculated again using the new values of $s$ and $q$, and including only the particles within the new ellipsoidal region. This iterative process can be repeated until the variance in both axis ratios, $s$ and $q$, is less than a given tolerance or up to a maximum number of iterations \citep[see][for more details]{Allgood2006}. In {\sc rockstar} the halo centres are not recalculated within the iterative procedure. In this work we derive the axis ratios using three different methods:

\begin{itemize}

\item a weighted inertia tensor in a spherical window (i.e. only one iteration) with axis ratios labelled as $s_{sph}$;
\item an iterative weighted inertia tensor (10 iterations) with axis ratios labelled as $s_{w}$;
\item an iterative unweighted or reduced inertia tensor (10 iterations) with axis ratios labelled as $s_{un}$.

\end{itemize}

\begin{figure*}
\includegraphics[width=5.8cm,angle=0.0]{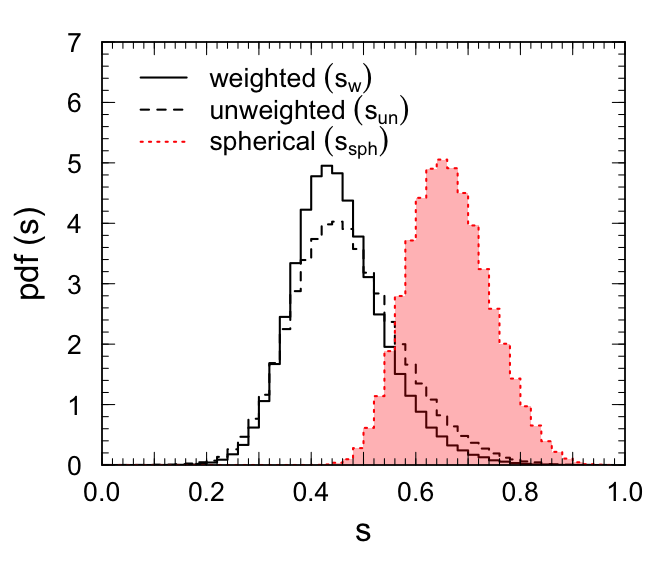}
\includegraphics[width=5.8cm,angle=0.0]{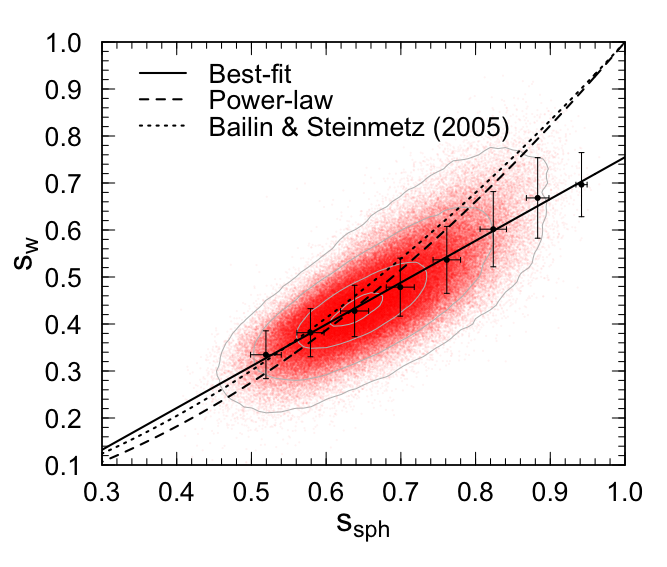}
\includegraphics[width=5.8cm,angle=0.0]{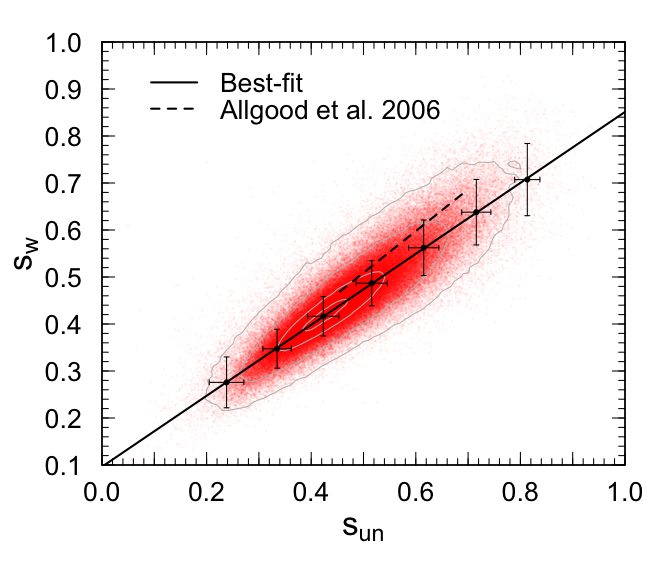}
\caption{Left-hand panel: distribution of virial axis ratios for $\textsc{hmd}$ relaxed haloes ($z=0$) obtained by different methods. Central panel: distribution of virial axis ratios in the plane $s_{w}$ versus $s_{sph}$. The black circles and error bars are the running means and 1$\sigma$ levels in eight bins in $s_{sph}$. The grey contours overlaid to the image show the intensity levels corresponding to 1, 10, 50 and 90 per cent of the probability peak. The dashed black line indicates the result of the fit with a power law to the data points ($s_{w} \simeq s^{1.86}_{sph}$), while red dotted line shows the findings of \citet{Bailin2005}. The solid black line corresponds to the linear best fit to the data points. Right-hand panel: distribution of virial axis ratios in the plane $s_{w}$ versus $s_{un}$. The black circles and error bars are the running means and 1$\sigma$ levels in eight bins in $s_{w}$. The solid black line corresponds to the linear best fit to the data points, while the dashed black line shows the results of \citet{Allgood2006} for all haloes (both relaxed and unrelaxed ones).}
\label{fig:weighted_contours}
\end{figure*}

In order to compare between the different methods, we computed the axis ratios considering the three methods described above for relaxed haloes only in the $\textsc{hmd}$ simulation at $z = 0.00$. In Fig.~\ref{fig:weighted_contours} we show the distribution of the different virial axis ratios. For a direct comparison we also show a scatter plot for the axis ratios $s_w$ versus $s_{sph}$. We perform a power-law fit to the data finding $s_{w} = s^{\gamma}_{sph}$ with $\gamma = {1.86\pm0.13}$ and $R^2 = 0.969$ (where $R^2$ is the coefficient of determination). A similar result was found in \citet{Bailin2005} with $\alpha = \sqrt{3} \approx 1.73$. However, our data points are best fit by a linear relation
\begin{equation} 
s_{w} = (-0.13\pm0.02) + (0.89\pm0.02) s_{sph}\;,
\label{eq:sw2sph}
\end{equation} 
with higher $R^2 = 0.996$ and $\sigma$ = 0.02. We also examined the difference in the determination of the virial axis ratios derived with the weighted ($s_w$) and the unweighted ($s_{un}$) iterative inertia tensor in Fig.~\ref{fig:weighted_contours} (right-hand panel). Data points are best fit by a linear relation as well
\begin{equation} 
s_{w} = (0.096\pm0.002) + (0.755\pm0.004) s_{un}\;,
\label{eq:sw2nw}
\end{equation} 
with a dispersion of $\sigma$ = 0.004. The weighted axis ratios are $\sim15$ per cent lower than those obtained with the unweighted inertia tensor. Although not plotted for clarity, we do not find any significant dependence on the halo mass in the above relations. For comparison we also show the results obtained by \citet{Allgood2006} with a slope of about 0.87. Although the slope is different than our findings ($\sim$0.75), we recall that \citet{Allgood2006} included the whole sample in the analysis (both relaxed and unrelaxed haloes). Besides, they measured the axis ratios at an inner radius ($R~\sim~$0.3$R_{\rm{vir}}$), while in the right-hand panel of Fig.~\ref{fig:weighted_contours} we determine the axis ratios at $R=R_{\rm{vir}}$, which can lead to differences in the measurement of the axis ratios, particularly for the unweighted inertia tensor.

Since the weighted inertia tensor weights the contribution from each particle in the sum by the distance to the particle squared, it is less sensitive to large substructure in the outer regions of the volume analysed. Moreover, the unweighted inertia tensor is more biased than the weighted inertia tensor due to particles at large radii. For these reasons, hereafter, we show the axis ratios derived from the weighted inertia tensor (i.e., $s = s_w$) for the four simulated volumes at $z = (0.00, 0.33, 0.66, 1.00)$.

\subsection{Minor-to-major axis ratio}
\label{sec:minortomajor}

As already proposed by many authors \citep{JS02, Allgood2006, Bett2007, Schneider2012}, the axis ratios depend on the mass and the redshift of the halo, however this dependence needs to be tested for the most massive haloes in large simulation boxes. Recently, \citet{Bonamigo2015} proposed an interesting procedure to account for the redshift dependence of the halo shapes in terms of the peak height $\nu = \delta_c / \sigma(M)$, where $\delta_c$ is the critical overdensity of the spherical collapse model and $\sigma(M)$ is the variance in the initial density field smoothed on a linear scale of a uniform sphere of mass $M$. Within this framework, the characteristic mass ($M_{*}$) is defined as the mass giving $\nu (z, M_{*}) = 1$ and therefore, dark matter haloes with $\nu > 1$ will have masses larger than the typical haloes collapsing at a given redshift. See Appendix \ref{sec:nu} for the details on how to compute the peak height $\nu$ for a halo of a given mass $M$.

\begin{figure}
\includegraphics[width=8.5cm,angle=0.0]{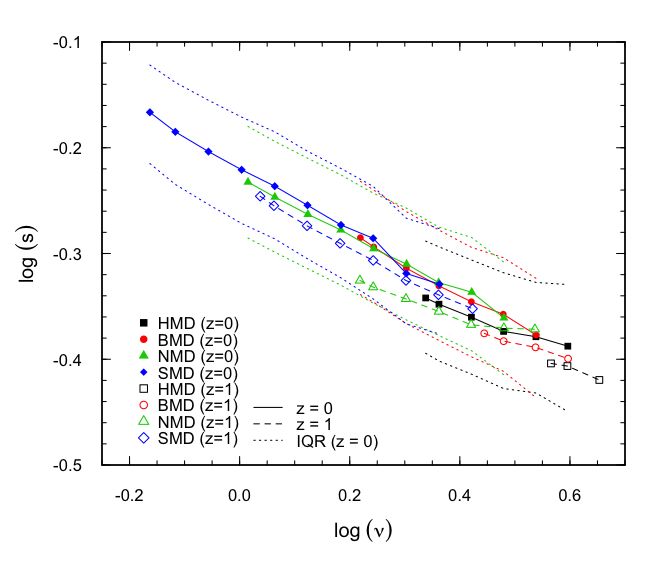}
\caption{Virial axis ratio $s$ as a function of the peak height in log--log space for the four simulated boxes. Different points correspond to the median values of $s$ in each bin considered. The dotted lines enclosed the Q1 and Q3 quartiles for the distribution of $s$ at $z=0$.}
\label{fig:logs_comparison}
\end{figure}

In Fig.~\ref{fig:logs_comparison} we show the axis ratio $s$ distributions as a function of the peak height $\nu$ of the halo for the four simulated box volumes (only relaxed haloes). Since the merger rate of massive haloes is larger, we expect them to show smaller axis ratio than low-mass haloes (with median $s \approx 0.4$), as already pointed out by \citet{JS02,Allgood2006,munozcuartas} and \citet{Despali2014}. Moreover, haloes at higher redshift are also more elongated, in agreement with the findings of \citet{munozcuartas} and \citet{Despali2014}. At the same time, axis ratios for a given halo mass are slightly smaller as the simulated box volume increases (mostly at $z=1$). This effect can be explained because the halo shape is noisier (and therefore more likely to be aspherical) for lower particle counts. Nevertheless, these differences are less than 20 per cent of the interquartile range of the relation (i.e. $\rm{IQR} = \rm{Q}_3 - \rm{Q}_1$).

It is also important to notice that using the same procedure as in \citet{Bonamigo2015}, we do not completely remove the redshift dependence on the axis ratio distributions (Fig.~\ref{fig:logs_comparison}). For this reason, we proposed a modified scaling relation between the peak height $\nu$ and the axis ratio $s$ in log--log space, which explicitly includes the redshift dependence as function of the linear growth rate $D(z)$ as follows:

\begin{equation} 
\mathrm{log}(s) = b + a ~ \mathrm{log}~(\nu^{\prime})\;,
\label{eq:relation_axial}
\end{equation} 
where we have defined $\nu^{\prime} = \nu ~ D(z)^c$. Since $D(0) = 1$ by definition, this scaling relation is identical to the one presented by \citet{Bonamigo2015} only at $z=0$ (or in the case of $c=0$). Therefore, the redshift dependence of  the scaling relation $s-\nu^{\prime}$ can be assessed considering that
\begin{equation} 
\nu^{\prime} (M, z) = \nu (M, z) ~ D(z)^c = \nu_0 (M) D(z)^{c-1}\;.
\label{eq:nu_redshift}
\end{equation} 

\begin{figure}
\includegraphics[width=8.5cm,angle=0.0]{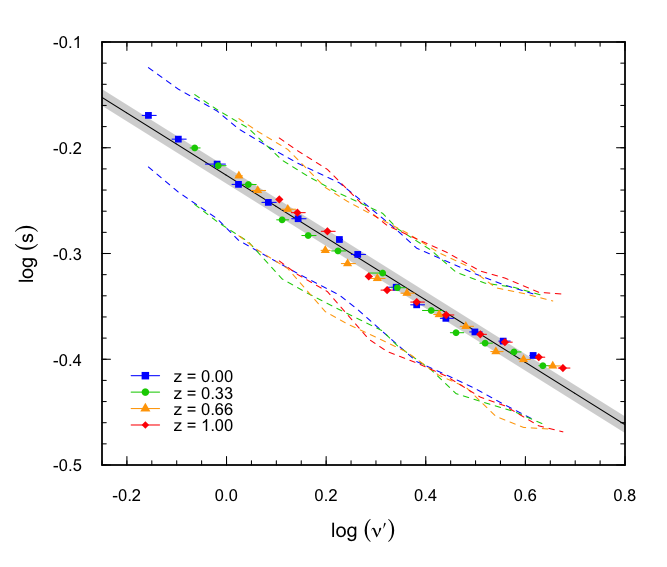}
\caption{Virial axis ratio $s$ as function of the peak height (expressed as $\nu^{\prime}$) in log--log space for relaxed haloes in the four simulated boxes. Different points correspond to the median values of the distribution at different redshifts, while error bars enclose the Q1 and Q3 quartiles of the peak height in each bin considered. Dashed lines show the Q1 and Q3 quartiles around the median values of $s$ in each bin of $\nu$. The solid black line shows the linear fit in log--log space to the data points using equation~(\ref{eq:relation_axial}), while the shade grey region corresponds to the standard deviation of the residuals (equation~\ref{eq:std}).}
\label{fig:logs_all}
\end{figure}

In Fig.~\ref{fig:logs_all} we show the median values for the axis ratio $s$ as a  function of $\nu^{\prime}$ for the relaxed haloes in the four simulated boxes. We binned the haloes according to their $\nu$ in log--log space and considering $\Delta \mathrm{log} ~ \nu = 0.06$. The error bars in the figure correspond to the 25 and 75 per cent probabilities of $\nu^{\prime}$ in each bin. The best fit by equation~(\ref{eq:relation_axial}) to the data points is obtained with $ac = 0.08 \pm 0.01$, which reveals a stronger dependence with redshift than found by \citet{Bonamigo2015}. The standard deviation of the residuals of the fit is $\sigma = 0.008$, defined as

\begin{equation} 
\sigma = \frac{1}{N_{\rm dof}} \sum_{i}R_i^2\;,
\label{eq:std}
\end{equation} 
where $R_i$ is the residual measured in the \textit{i}th bin and $N_{\rm{dof}}$ is the number of degrees of freedom (i.e. the number of bins minus the number of free parameters in the fit). The procedure used to perform the fit is a general-purpose optimization based on \cite{Nelder01011965} algorithm.

The best fitting parameters are shown in Table~\ref{tb:logs_lognu_fit} for relaxed haloes, unrelaxed haloes and the whole sample. We also show the results for the axis ratio measured at $\Delta = 500\rho_c$ as function of the corresponding peak height ($\nu_{500}$) for a halo of a given mass $M_{\rm{500}}$ at redshift $z$.

\begin{table*}
\begin{center}
\caption{Best-fitting parameters to the scaling relation between $\nu$ and axis ratio (equation~\ref{eq:relation_axial}). First column indicates the selection criteria (rlx = relaxed haloes, unr = unrelaxed haloes, all = full sample); second column shows the definition of virialization adopted; columns 3,4 and 5 correspond to the $b$, $a$ and $c$ parameters; column 6 shows the product of columns 4 and 5; last column shows the standard deviation of the residuals.}
\label{tb:logs_lognu_fit}
\begin{tabular}{ccccccc}
\hline
$\rm{Select.}$ & $\Delta$ & $b$ & $a$ & $c$ & $ac$ & $\sigma$\\
\hline
rlx & $\Delta_{\rm{vir}}~\rho_{b}$ & -0.226 $\pm$ 0.002 & -0.295 $\pm$ 0.005 & -0.27 $\pm$ 0.05 & 0.08 $\pm$ 0.01 & 0.008 \\
unr & $\Delta_{\rm{vir}}~\rho_{b}$ & -0.333 $\pm$ 0.002 & -0.227 $\pm$ 0.004 & -0.11 $\pm$ 0.04 & 0.03 $\pm$ 0.01 & 0.005 \\
all & $\Delta_{\rm{vir}}~\rho_{b}$ & -0.246 $\pm$ 0.002 & -0.313 $\pm$ 0.005 & -0.31 $\pm$ 0.05 & 0.09 $\pm$ 0.02 & 0.008 \\
rlx & $500~\rho_{c}$ & -0.270 $\pm$ 0.003 & -0.303 $\pm$ 0.009 & -0.13 $\pm$ 0.07 & 0.04 $\pm$ 0.02 & 0.01 \\
unr & $500~\rho_{c}$ & -0.301 $\pm$ 0.003 & -0.329 $\pm$ 0.009 & -0.11 $\pm$ 0.07 & 0.03 $\pm$ 0.02 & 0.01 \\
all & $500~\rho_{c}$ & -0.279 $\pm$ 0.003 & -0.310 $\pm$ 0.007 & -0.21 $\pm$ 0.06 & 0.07 $\pm$ 0.02 & 0.01 \\
\end{tabular}
\end{center}
\end{table*}

Given the linear dependence (in log scale)  of the axis ratio as a function of the peak height ($\nu^{\prime}$), we derived the scaled axis ratios  in a similar way as performed by previous authors: $
	\log\left(s\right) = a \log\left(\nu^{\prime}\right) + b 
	\Rightarrow \tilde{s} = 10^b = 10^{\log s -a \log\left(\nu^{\prime}\right)} = s \, (\nu^{\prime})^{-a}.$ The black line in Fig.~\ref{fig:logs_all} corresponds to the best fit to the median axis ratios for relaxed haloes with slope $a = -0.295$. The \textit{y}-intercept point  is the logarithm of the median axis ratio at $\nu^{\prime} = 1$, which corresponds to $M = M_{*}$ at $z=0$. Thus, the fit yields to $\tilde{s} (z=0, M_{*})= 10^\textit{b} \approx 0.59$. 

\begin{figure}
\includegraphics[width=8.5cm,angle=0.0]{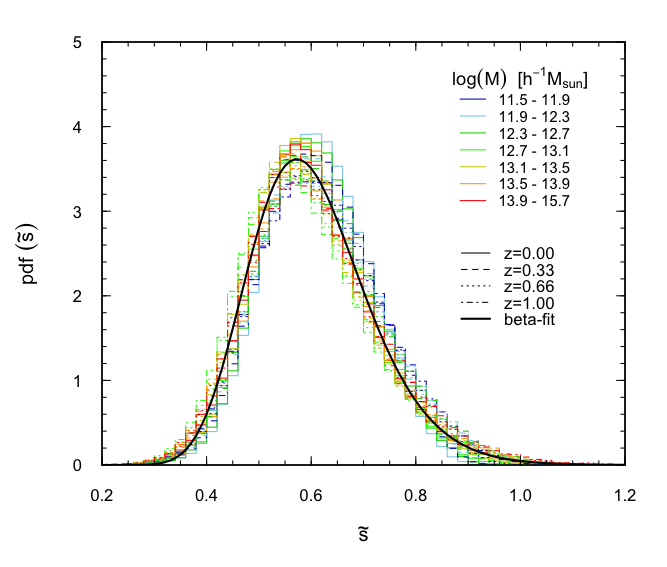}
\caption{Distribution of the scaled axis ratio $\tilde{s}$ in different mass bins for relaxed haloes in the four simulated boxes. Different line types indicate the different redshifts. The black line corresponds to the best fit to a lognormal distribution (equation~\ref{eq:log-normal_dist}).}
\label{fig:scaled_hmd}
\end{figure}

The scaled axis ratio distribution $\tilde{s}$ of relaxed haloes shows a fairly universal distribution almost independently on the mass and redshift of the haloes for the four simulated boxes (Fig.~\ref{fig:scaled_hmd}). The probability distribution function of the scaled axis ratio ($\tilde{s}$) is well fitted by a lognormal distribution as proposed by \citet{Bonamigo2015}:
\begin{equation}
	p(\tilde{s},\mu,\sigma) = \frac{1}{\tilde{s} \sqrt{2\pi}\sigma} \exp\left( -\frac{\left( \ln \tilde{s} - \mu \right)^2}{2\sigma^2} \right)
\label{eq:log-normal_dist}
\end{equation}
The parameters of the fitted function are $\mu$  = --0.52 and $\sigma$ = 0.19, which can be converted to the median and the standard deviation of the Gaussian distribution:
\begin{equation}
\begin{split}
	\text{median }& = e^\mu = 0.59,\\
	\text{std }& = \sqrt{(e^{\sigma^2} - 1 ) e^{2\mu + \sigma^2} } = 0.11\;.
\end{split}
\label{eq:hmd_lognormal_parameters}
\end{equation}

The results for the scaled axis ratio of relaxed haloes are in good agreement with those obtained by \citet{Bonamigo2015}, which found a median of 0.61 and a standard deviation of 0.13. 

In the case of axis ratios measured at $500 \rho_{c}$ (see Table~\ref{tb:logs_lognu_fit}), the best-fitting parameters show a lower normalization than for $\Delta_{\rm{vir}} \rho_{b}$. In particular, for relaxed haloes $\tilde{s}_{500} (z=0, M_{*})= 10^\textit{b}  \approx 0.54$. The redshift dependence is also lower for the axis ratios measured at $500 \rho_{c}$ with $c=-0.13$. The parameters of the log-normal distribution for $\tilde{s}_{500}$ are $\mu = -0.62$  and $\sigma = 0.19$ ($\text{median} = 0.54$ and $\text{std} = 0.10$, respectively).


Therefore,  we can now study the distribution of axis ratios for any halo of a given mass (i.e. peak height)  and redshift. First we generate one random value of the scaled axis from the log-normal distributions, either for $\tilde{s}$ or  $\tilde{s}_{500}$. Then, by inverting the rescaling relation, $s = \tilde{s}~\nu^{\prime a}$, it is possible to infer the axis ratio of the halo given its peak height $\nu(z,M)=\nu^\prime~D(z)^{-c}$. Then it can then be converted in mass for a given redshift of the halo and for a particular cosmology (see Appendix \ref{sec:nu}).

\subsection{Intermediate-to-major axis ratio}
\label{sec:middle}

\begin{figure*}
 \includegraphics[width=8.5cm,angle=0.0]{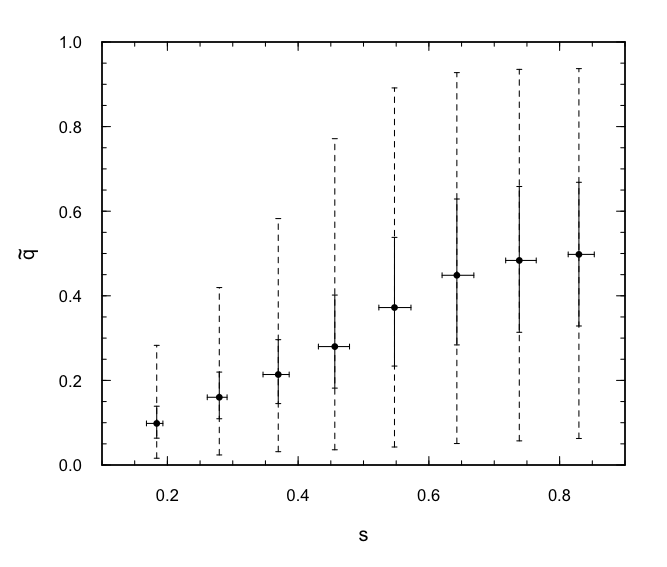}
 \includegraphics[width=8.5cm,angle=0.0]{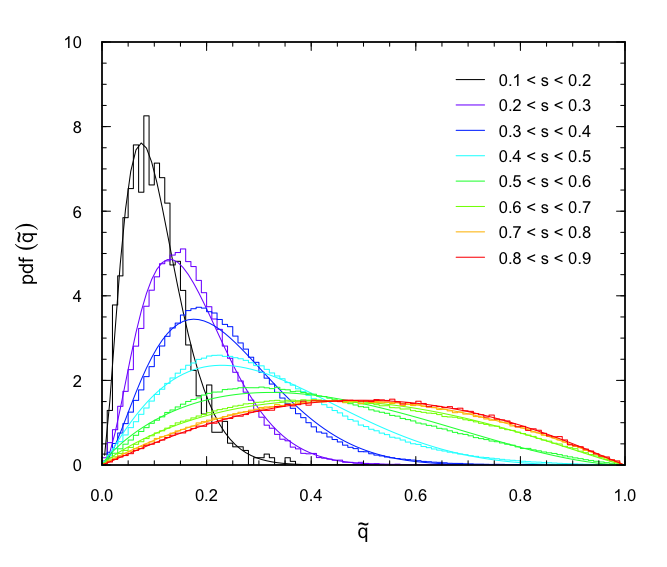}
  \caption{Left-hand panel: distribution of $\tilde{q} = (q-s)/(1-s)$ as a function of the virial axis ratio ($s$) for relaxed haloes. Points indicate the median values of $s$ and $\tilde{q}$, while solid (dashed) error bars represent the 25--75 per cent (1--99 per cent) probabilities of the distribution. Right-hand panel: probability distributions of $\tilde{q}$ for different values of $s$ (histograms) and the fitting function to a $\beta$ distribution (lines) for relaxed haloes. Both panels include all the redshifts analysed in this work.
 \label{fig:minor-to-major}}
\end{figure*}

\begin{figure}
\includegraphics[width=8.5cm,angle=0.0]{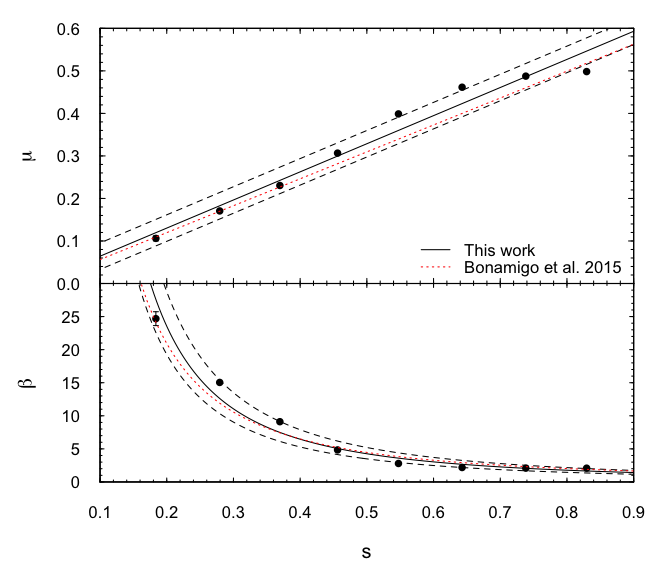}
\caption{Best-fitting parameters of equation~(\ref{eq:beta_function}) for different values of the virial axis ratio ($s$) for relaxed haloes (solid points). Solid lines show the best fit to the data points given by equation~(\ref{eq:hmd_beta_parameters}), while dashed lines correspond to the standard deviation of the residuals. For comparison the results of \citet{Bonamigo2015} are shown as dotted lines.}
\label{fig:beta_function}
\end{figure}

As already proposed by several authors \citep{Schneider2012, Bonamigo2015}, the minor-to-major axis ratio is better described by the rescaled quantity, $\tilde{q} = (q-s)/(1-s)$,  that eliminates the limited interval at which $q$ is defined by construction ($q \geqslant s$ and $q < 1$). Fig~\ref{fig:minor-to-major} shows the relation between the rescaled axis ratio $\tilde{q}$ and $s$ for relaxed haloes. We computed the probability distributions of $\tilde{q}$ in bins of different $s$.  We have  fitted the probability distribution of $\tilde{q}$ in each bin of $s$ to a $\beta$-function  \citep{Bonamigo2015}
\begin{equation}
	p(\tilde{q},\alpha,\beta) = \frac{1}{B(\alpha,\beta)} \tilde{q}^{\alpha-1}(1-\tilde{q})^{\beta-1}\;,
\label{eq:beta_function}
\end{equation}
with two shape parameters $\alpha$ and $\beta$ and the normalization factor $1/B(\alpha,\beta)$. The mean value of the $\beta$ distribution is defined as $\mu = 1/(1+\beta / \alpha)$. From the fitting procedure, we derived $\mu$ and $\beta$ parameters for each bin in $s$. Fig.~\ref{fig:beta_function} shows the dependence of these parameters on the axis ratio $s$. The solid black lines in the figure correspond to the best fit of these two parameters as follows:
\begin{equation}
\begin{split}
	\mu (s) = -0.002 +0.661s\\
	\beta (s) = 1.170s^{-1.865}\;.
\end{split}
\label{eq:hmd_beta_parameters}
\end{equation}
These results  are in fairly  good agreement with those reported by  \citet{Bonamigo2015}. Nevertheless  we   find  a slightly higher mean value of the $\beta$ distribution for low values of $s$. Given these two equations and $p(\tilde{q} | s)$, it is possible to infer the distribution of $\tilde{q}$ for  a given $s$ along with its scatter.

Additionally, we have repeated this calculation considering the axis ratios at an overdensity $500 \rho_c$ (denoted as $s_{500}$ and $\tilde{q}_{500}$). The probability distributions of $\tilde{q}_{500}$ in different bins of $s_{500}$ are also well described by a $\beta$ distribution with parameters $\mu_{500}$ and $\beta_{500}$. The best fit of these two parameters are
\begin{equation}
\begin{split}
	\mu_{500} (s_{500}) = -0.014 + 0.676 s_{500}\\
	\beta_{500} (s_{500}) = 1.159 s^{-1.792}_{500}\;.
\end{split}
\label{eq:hmd_beta500_parameters}
\end{equation}

\subsection{Triaxiality and major axis alignment}
\label{sec:align}

\begin{figure}
\begin{center}
 \includegraphics[width=8.5cm,angle=0.0]{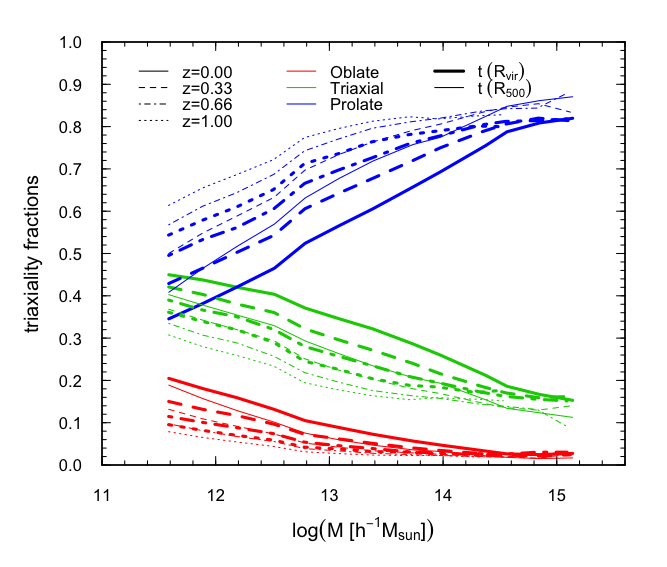}
\end{center}
 \caption{Fraction of \textit{oblate} (red lines), \textit{triaxial} (green lines) and \textit{prolate} (blue lines) haloes as a function of mass for relaxed haloes according to their triaxiality parameter $t$ (equation~\ref{eq:triax}). Thicker lines show to the triaxiality parameter measured at $\rm{R_{vir}}$, while thinner lines show to the same parameter measured at $\rm{R_{500}}$. Different line styles (solid, dashed, dashed-dotted, dotted) correspond to the different redshift analysed in this work.}
\label{fig:triaxiality}
\end{figure}

The triaxiality of an ellipsoid can be expressed as  \citep{Franx1991}:

\begin{equation}
t \equiv \frac{a^2 - b^2}{a^2 - c^2} = \frac{1 -q^2}{1 - s^2}.
\label{eq:triax}
\end{equation}
Depending on the value of the parameter $t$, an ellipsoid is considered as: \textit{oblate} if $0 < t < 1/3$; \textit{triaxial} if $1/3 < t < 2/3$ and \textit{prolate} if $2/3 < t < 1$. In Fig.~\ref{fig:triaxiality} we show the fraction of haloes classified in this way as a function of the halo mass (only relaxed haloes). haloes show preferentially a prolate shape with fraction larger than the 45 per cent for haloes with masses above 10$^{12}~h^{-1} \rm{M}_\odot$ at $z=0.00$. At higher redshifts, the fraction of prolate haloes increase, while the fraction of triaxial and oblate decrease. The more massive the haloes are, the larger the fraction of prolate haloes. Consequently, the opposite is true for the triaxial and oblate fraction. In particular, for haloes at $z=0.00$ with masses larger than 10$^{14}~h^{-1} \rm{M}_\odot$ are dominated by a prolate shape (with fraction above 70 per cent), while only a few massive haloes are found to be triaxial or oblate in shape (with fraction below 20 and 5 per cent, respectively). At higher redshift, the fraction of prolate massive haloes are even larger, with values of about 80 per cent. This is in support of the idea that haloes with peak height $\nu > 1$ are expected to be undergoing a higher rate of merging processes than haloes with $\nu < 1$, and that the mergers happen along preferred directions \citep{Knebe2004,Faltenbacher2005,Zentner2005}.

Besides, the fraction of prolate haloes at inner radius ($\rm{R_{500}}$) are larger on average than those measured at $\rm{R_{vir}}$, while the fractions of triaxial and oblate haloes are slightly lower towards the centre, independently of the halo mass. About 5--10 per cent of the haloes classified as triaxial attending to the triaxiality measured at $\rm{R_{vir}}$, would be classified as prolate when looking at inner radius. In the case of oblate fraction these differences are below the 2 per cent.

\begin{figure}
\includegraphics[width=8.5cm,angle=0.0]{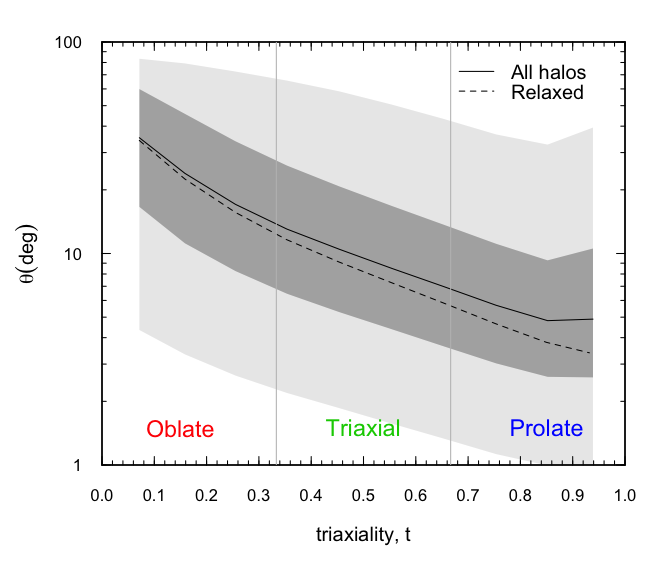}
\caption{Distribution of the major axis alignment ($\theta$, in degrees) as a function of the triaxiality parameter $t$ for the whole sample at $z=0.00$. Solid and dashed black lines correspond to the median values of the distribution for the whole and relaxed samples, respectively. Dark grey shaded region indicates the Q1 and Q3 quartiles around the median values of $\theta$, while grey shaded region indicates the 1 and 99 per cent probabilities of the distribution of $\theta$ for the whole sample only.}
\label{fig:align}
\end{figure}

Since we measured the axis ratios at two different overdensities, we examine the angle between the major axis at $\Delta_{\rm{vir}} \rho_{b}$ and $500 \rho_{c}$, define here as $\theta$. For the given cosmology, the radius measured at $500 \rho_{c}$, denoted as $\rm{R_{500}}$, roughly corresponds to $0.5 \rm{R_{vir}}$. In Fig.~\ref{fig:align} we show the alignment of the major axis as a function of the axis ratio ($s$). We find that the major axis aligns pretty well for prolate haloes, with 99(75) per cent of then below $\theta < 40º(11º)$ (Fig.~\ref{fig:align}). Since the most massive haloes are preferentially prolate (Fig.~\ref{fig:triaxiality}), they appear to be well aligned, but this would be simply because less massive haloes are more spherical and thus the direction of the major axis is less accurate than in the case of cluster-sized haloes. On the other hand, triaxial and oblate haloes show larger misalignments, with $\theta \gtrsim$10º (on average). The angle between the major axis increases to $\theta \simeq$ 12º--13º (on average) for haloes $t \lesssim 0.2$. We also find a very small fraction of haloes haloes with misalignments close to $\theta = 90º$ and low triaxiality. These large misalignments can be due to less massive haloes with larger $q$ (thus $b \approx a$), for which the direction of the major axis is arbitrary within a plane. Finally, relaxed haloes show, on average, slightly lower misalignments with respect to the whole sample (less than $2º$ differences in the case of prolate haloes). 

\subsection{Distribution of the minima axis ratio $s$}
\label{sec:minima}

As already mentioned, triaxiality of dark matter haloes has a strong impact on the lensing cross-sections. Since massive haloes are more prolate than less massive ones, this effect is even stronger in the case of clusters of galaxies. In particular, \citet{Waizmann12} found that the impact of the tail of the axis ratio distribution on the largest Einstein radii distribution is substantial. However, they also argued that due to the limited knowledge of the statistics of extremely small axis ratios, it is not possible to clearly define a proper choice of the cut-off (if present) until the triaxiality distributions of large halo samples (up to cluster masses) are studied in numerical simulations. The results presented in this work provide scaling relations to characterize the halo triaxiality that are derived from a huge sample of haloes (including thousands of galaxy clusters). 

\begin{figure}
\includegraphics[width=8.5cm,angle=0.0]{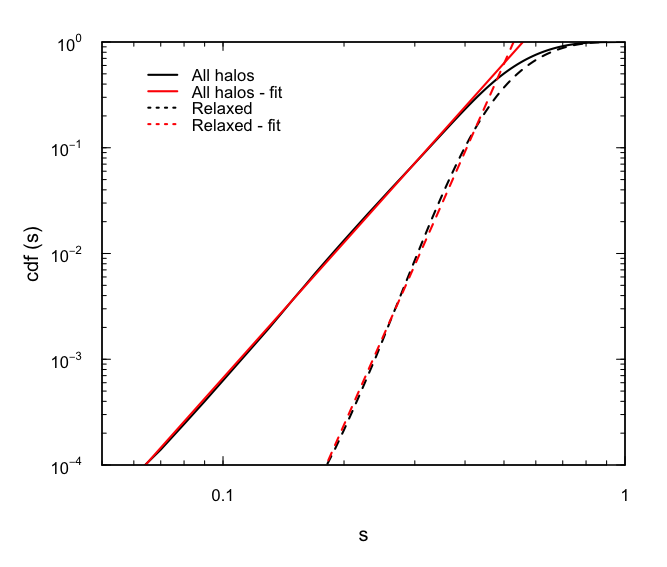}
\caption{Cumulative distribution of the minima axis ratio, $P(s \leqslant S)$, for the whole sample (solid black line) and for relaxed haloes (dashed black line), including all the redshifts analysed in this work. Red solid and dashed lines correspond to the best fitted to the whole and relaxed distributions (equations~\ref{eq:fit_minima_all} and \ref{eq:fit_minima_rlx}, respectively).}
\label{fig:minima}
\end{figure}

Since new  semi-analytic lensing models could be implemented with our prescriptions on the shape of dark matter haloes, we consider important to check that extremely elongated haloes do not propagate through the sampling procedure. For this reason and in order to clarify whether a cut-off in the minor-to-major axis ratio distribution is needed or not, we present in this section the distribution of the minima of the minor-to-major axis ratio. In this analysis we include all the haloes (both relaxed and unrelaxed) in the four simulated boxes at the four redshifts analysed. The cumulative distribution of $s$, denoted as $P(s<S)$, is shown in Fig.~\ref{fig:minima}. As expected, unrelaxed haloes populate the lower tail of the distribution of axis ratios. The probability of finding a halo with $s < 0.1$ is $P(s<0.1) =10^{-3.20} \approx 6 \times 10^{-4}$, while for relaxed haloes this probability goes down to $P(s<0.1) \approx 10^{-5.62}$ (about two haloes per million). In particular, we find only 13 relaxed haloes with $s<0.1$ from a total of almost 5,5 million relaxed haloes (all redshifts included). The relaxed halo with the minimum axis ratio in our sample corresponds to $s=0.07$. For the whole sample (composed of more than 8,6 million haloes) we find 5,430 haloes with $s<0.1$.

Given the linear shape of the distribution in log--log space, we give in this section explicit relations for $P(s<S)$ as a function of $S$ that are valid for $S\lesssim0.4$ for the whole sample

\begin{equation}
	\mathrm{log} \left[P(s<S)\right]  = (1.08\pm0.03) + (4.26\pm0.03)~\mathrm{log}(S),\;
\label{eq:fit_minima_all}
\end{equation}
and for relaxed haloes
\begin{equation}
	\mathrm{log} \left[P(s<S)\right] = (2.38\pm0.08) + (8.58\pm0.12)~\mathrm{log}(S)\;.
\label{eq:fit_minima_rlx}
\end{equation}

Although not plotted in Fig.~\ref{fig:minima}, we do not find any significant dependence with redshift on the $P(s<S)$ distribution.

\section{Comparison with previous results}
\label{sec:comparison}

\begin{figure*}
 \includegraphics[width=0.7\hsize,angle=0.0]{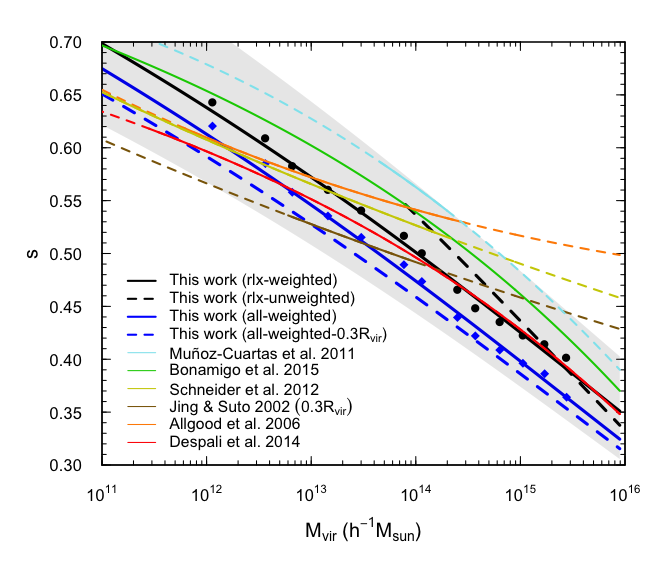}
 \caption{Comparison of the mass-shape scaling relation with previous studies at $z=0$. Black and dark-blue lines show the results of this work using the weighted inertia tensor at $\rm{R_{vir}}$ for relaxed haloes and the whole sample, respectively. The grey shaded region indicates the IQR associated with the black line. The dashed black line shows the axis ratio derived using the unweighted inertia tensor at $\rm{R_{vir}}$ (only for $\textsc{hmd}$ simulation), while the dashed dark-blue line corresponds to the axis ratio measured at 0.3$\rm{R_{vir}}$ for the whole sample (weighted inertia tensor). Black filled circles and blue filled diamonds correspond to the median axis ratio derived at $\rm{R_{vir}}$ using the weighted inertia tensor for relaxed haloes and the whole sample, respectively. Thin solid lines show the results from other studies in the mass range analysed in each case, while extrapolation outside the mass range is denoted by thin dashed lines.
\label{fig:comparison}}
\end{figure*}

In this section we present a comparison of the mass-shape results presented  in this work with  previous estimations from literature. In Fig.~\ref{fig:comparison} we show the median minor-to-major axis ratios ($s$) as a function of the halo mass from the four simulated boxes at $z=0$. We display separately the axis ratio for relaxed haloes (black filled circles) and for the whole sample (blue filled diamonds) obtained with the weighted inertia tensor, $s_w$. The dashed black line indicates the scaling relation for the unweighted inertia tensor using equation~(\ref{eq:sw2nw}). In order to show the scatter of the mass-shape relation, we also plot the IQR corresponding to the median axis ratios for the relaxed sample and using the weighted inertia tensor (shaded grey region). For clarity we do not include that value for the rest of the scaling relation, although their IQR is very similar to that shown in the figure. Results from other authors are shown with thinner lines, the solid part of the curves represent the mass limits studied in each case.

As it can be seen in Fig.~\ref{fig:comparison}, there is an overall agreement in the minor-to-major axis ratio ($s$) as a function of halo mass, with more massive haloes showing less spherical shapes. The scatter in the axis ratio relations, which is constrained within $\sim$10 per cent, is mainly due to the differences in the method of measuring the shapes (Section~\ref{sec:shapes}) and the definition of halo masses and the axis radius (i.e. the overdensity at which it is measured).  For a reliable comparison, we have converted the results from other authors to redshift $z=0$ and the cosmology assumed in this work. Therefore, the differences shown in Fig.~\ref{fig:comparison} cannot be attributed to differences in the cosmology assumed in each work. Nevertheless, we would like to remark that the MultiDark suite constitutes a coherent and homogeneous data set of cosmological simulations (performed with {\sc l-gadget}-2 code for identical cosmological parameters), for which all the haloes are identified using the {\sc rockstar} halo finder. In particular, the $\textsc{hmd}$ simulation (4$~h^{-1}$Gpc cube side) allows us to study in detail the shape of simulated dark matter haloes in the cluster mass range with unprecedented statistics and resolution (more than $8\times10^5$ haloes at $z=0$ with more than 3,000 particles per halo). As we described in Section \ref{sec:catalogs}, given that massive haloes are more aspherical, larger simulated volumes show (on average) larger unrelaxed fraction than smaller ones ($\sim$40 per cent for the $\textsc{hmd}$ simulation, while $\sim$20 per cent for the $\textsc{smd}$ simulation). The fraction of unrelaxed haloes found in the simulations have a significant impact on the mass-shape scaling relations when all the haloes are included in the analysis (i.e. not selected by their relaxation state), as we have shown in Section~\ref{sec:minortomajor}. On the other hand, small simulated volumes will not be strongly affected by this selection criteria because of the small fraction of unrelaxed haloes found in them. In order to facilitate the comparison we list the information and details of previous studies on the shapes of dark matter haloes in Table~\ref{tb:comparison}.

The definition of the halo masses can be obviously a source of the scatter in the mass-shape scaling relations. In this work we used the definition given by the spherical collapse model (i.e. the mass inside the radius encompassing a given density contrast), then the procedure used to compute the axis ratio iteratively keeps the major axis equal to the original sphere radius. According to this description the ellipsoid is, by definition, inside the sphere and therefore its mass is always smaller than the mass of the initial sphere. Nevertheless, \citet{munozcuartas} checked that these differences are below 20 per cent. Alternatively, in \citet{Bonamigo2015} and \citet{Despali2014} the halo mass is defined as the mass within the ellipsoid that encloses a given overdensity. Therefore, the best fitting ellipsoid always contains a mass larger or equal to that of the original sphere. \citet{Despali2013} found that elliptical masses are on average about 3 per cent larger than spherical ones for cluster-size haloes and 5 per cent larger for lower mass haloes. Taking all this into account, if spherical masses showed in Fig.~\ref{fig:comparison} would be converted into elliptical masses, the results slightly shifted to the right (i.e. towards higher masses), leading to a better agreement with \citet{Bonamigo2015}.

Additionally, since ellipsoid used in this work to compute the axis ratio is inside the original sphere, we measure an inner shell of the halo mass distribution, which we know may be on average more aspherical towards the halo centre. For the same reason, the method used in \citet{Bonamigo2015} to determine the axis ratios could lead to more spherical shapes as it extends the computation to radii beyond the original sphere. 

Given that the peak height is defined in terms of the critical overdensity of the spherical collapse model ($\delta_c$), we consider more appropriate to derive the peak height from the spherical halo mass. Besides, the mass function of dark matter haloes is generally described according to the spherical overdensity criteria, so that the virial halo masses obtained from the mass function are spherical rather than elliptical. For all these reasons, we consider the spherical definition of the virial halo mass more useful for semi-analytical lensing studies that infer the lensing properties of dark matter haloes sampled from a particular mass function.

The various definitions of the halo centre can also introduced some scatter in the mass-shape scaling relations. In this work we define the halo centres as the densest point in phase space, which are not recalculated within the iterative procedure. Therefore, if there is an offset between the centre of mass and the density peak, this contributes to the inferred shape. Since unrelaxed clusters tend to have larger offsets, this effect is stronger for studies including all haloes in their analysis. However, we do not expect the mass-shape scaling relations to be very sensitive to the choice of the halo centres if they are defined as the densest point or the most bounded particle. In other words, even in the case of unrelaxed haloes, the position of the densest point (or most bounded particle) will remain almost constant, while the centre of mass will change in position.

\citet{JS02} measured the axis ratios using particles within the isodensity surface of $2500 \rho_c$, which roughly corresponds to $0.3\rm{R_{vir}}$. Therefore, for a better comparison these results should be compared with our results for $s_{500}$ (dashed blue line). There is a good agreement on average for low-mass haloes, but given that our relation is steeper more massive haloes are less aspherical than found by \citet{JS02}. It should be notice that \citet{JS02} derived the shape properties of dark matter haloes extracted from five realizations of 100$~h^{-1}$Mpc box size and from 12 haloes with $10^6$ particles within the virial radius, and therefore they studied a narrower mass range. 

In the case of \citet{Allgood2006}, where analytical expressions on the radial dependence of the axis ratios are explicitly given, we convert their axis ratio relation from $0.3\rm{R_{vir}}$ to $\rm{R_{vir}}$. Although we find a steeper slope than \citet{Allgood2006}, the results are in agreement within errors for low-mass haloes when unrelaxed haloes are also included in the analysis (solid blue line).

Both \citet{Allgood2006} and \citet{Schneider2012} used the method based on the weighted inertia tensor. In particular, \citet{Schneider2012} derived the axis ratios measured at the virial radius, which allows a meaningful comparison with our results for relaxed haloes (solid black line). On the other hand, \citet{munozcuartas} and \citet{Bonamigo2015} studied the shapes using the unweighted inertia tensor. Although the results are in good agreement when comparing with our results for relaxed haloes using the unweighted inertia tensor (dashed black line), we obtain a lower normalization. In the case of \citet{munozcuartas}, shapes are obtained from a small sample of haloes extracted from small simulated volumes, less than 12,000 with masses above $5\times 10^{12}~h^{-1}$M$_\odot$ (i.e. $N_{\rm{p}} > 500$). For the mass-shape scaling relation, the authors restricted the analysis to an even smaller sample, including only haloes with $N_{\rm{p}} > 4000$.

On the other hand, for massive clusters we find axis ratios lower than those presented by \citet{Bonamigo2015} in about 5--8 per cent. It should be noticed that most of the haloes analysed in \citet{Bonamigo2015} are extracted from the Millennium XXL simulation, which is performed with 1-year \textit{Wilkinson Microwave Anisotropy Probe} (\textit{WMAP}1) cosmological parameters ($\Omega_{\rm{M}} = 0.25$, $\Omega_{\rm{b}} = 0.045$, $\Omega_{\Lambda} = 0.75$, $\sigma_8 = 0.9$, $h = 0.73$). Nevertheless, the authors scaled their results in order to account for the differences in the cosmological model. Despite the fact of the different mass resolution and box volume in the simulations, the main difference between the results of this work and the results presented by \citet{Bonamigo2015} is the definition of the halo mass, the method to determine the axis ratios (weighted and unweighted, respectively) and the criteria considered to select relaxed haloes. As we mentioned above, since massive haloes are more aspherical than lower mass haloes, we expect the differences between the spherical and the elliptical masses would slightly shift our mass-shape scaling relations to the right (i.e. towards higher mass), and hence reducing the observed scatter. Although we use the scaling relation given by equation~(\ref{eq:sw2nw}) to derive the unweighted shape-mass scaling relation (dashed black line), there is a better agreement with \citet{Bonamigo2015}. It is important to notice the different criteria adopted to select relaxed haloes: \citet{Bonamigo2015} fixed the offset between the most bounded particle and the centre of mass of the particles enclosed by the ellipsoid to be less than 5 per cent of their virial radius, while in this work we assume a more flexible value of 7 per cent. Therefore, the relaxed sample of haloes analysed in \citet{Bonamigo2015} included less distorted haloes than those incorporated in our analysis. For that reason, we expect that the shape-mass relation of \citet{Bonamigo2015} lies above our results (i.e. higher values of s) since unrelaxed haloes tend to be more elongated. In particular, this effect is more important for more massive haloes given that unrelaxed fraction are higher for higher halo masses, as it can be clearly seen in Fig.~\ref{fig:relaxed_fractions}.

Finally, the findings of \citet{Despali2014} are obtained from a set of simulations performed with almost the same cosmological parameters as the simulations used in this work, and based on the unweighted inertia tensor. We find a very good agreement with their predictions for relaxed haloes, even though they included in the analysis all the haloes regardless of their state of relaxation. Nevertheless, given that their simulated volumes are smaller than the ones analysed in this work (and therefore their fraction of unrelaxed haloes are also smaller), we do not expect a strong impact on the distribution of the median axis ratios when including the unrelaxed haloes. Again in this case, the scatter in the scaling relation can be due to the definition of the halo mass (either spherical or elliptical) and the used of the unweighted inertia tensor method to measure the axis ratios.

It is clear that there is a general agreement between the studies performed over a large range in halo mass, such as the results presented in this work and in \citet{Despali2014} and \citet{Bonamigo2015}. Besides, these mass--shape scaling relations are obtained in terms of the peak height $\nu$ that leads to steeper relations than those found in rest of the studies included in the comparison (which are derived in terms of the characteristic mass, $M_{\ast}$). Despite the different simulations and the various methodologies used to determine the halo shapes in the mentioned studies, the mass-shape relations show a similar slope but with a different ($\lesssim$10 per cent) normalization.

\begin{table*}
\begin{center}
\caption{Details of the simulations and the methods used to derive the mass-shape scaling relations from previous studies (shown in Fig.~\ref{fig:comparison}). First column indicates the referenced paper; second column shows the $\Lambda$CDM cosmological model considered; third column shows the size limits of the simulated boxes analysed (in $~h^{-1}$Gpc); column 4 corresponds mass limits per particle, $m_{\rm{p}}$ (in units $10^{10}~h^{-1}$M$_\odot$); column 5 shows the minimum number of particles per halo ($N_{\rm{p}}$); column 6 indicates the selection criteria (rlx = relaxed haloes, all = full sample); column 7 indicates the inertia tensor used (w = weighted, unw = unweighted); column 8 corresponds to the overdensity at which the axis ratios are measured.}
\label{tb:comparison}
\begin{tabular}{cccccccc}
\hline
$\rm{Reference}$ & $\textit{Cosmo.}^{a}$ & Box & $m_{\rm{p}}$ & $N_{\rm{p}}$ & $\rm{Select.}$ & $\rm{Tensor}$ & $\rm{\Delta}$\\
\hline
\rm{This work} & $Planck$ & $\left[0.4-4.0\right]$ & $\left[0.01-7.9\right]$ & $3000$ & $\rm{rlx, all}$ & $\rm{w, unw}$ & $\Delta_{\rm{vir}}\rho_{b},~500\rho_{c}$\\
\citet{Bonamigo2015} & $WMAP1$, $Planck$ & $\left[0.06-3.0\right]$ & $\left[0.002-0.6\right]$ & $1000$ & $\rm{rlx}$ & $\rm{unw}$ & $\Delta_{\rm{vir}}\rho_{b}$\\
\citet{Despali2014} & $Planck$ & $\left[0.1-2.0\right]$ & $\left[0.06-63\right]$ & $200$ & $\rm{all}$ & $\rm{unw}$ & $\Delta_{\rm{vir}}\rho_{b}$\\
\citet{Schneider2012} & $WMAP3$ & $\left[0.1-0.5\right]$ & $\left[<10^{-3}-0.09\right]$ & $500$ & $\rm{rlx}$ & $\rm{w}$ & $\Delta_{\rm{vir}}\rho_{b}$\\
\citet{munozcuartas} & $WMAP5$ & $\left[0.01-0.15\right]$ & $\left[0.001-1.13\right]$ & $4000$ & $\rm{rlx}$ & $\rm{unw}$ & $\Delta_{\rm{vir}}\rho_{b}$\\
\citet{Allgood2006} & $\Lambda$CDM & $\left[0.08-0.2\right]$ & $\left[0.03-4.0\right]$ & $7000$ & $\rm{all}$ & $\rm{w}$ & $\Delta_{\rm{vir}}\rho_{b}$$^{b}$\\
\citet{JS02}$^{c}$ & $\Lambda$CDM & $0.1$ & $0.06$ & $10000$ & $\rm{all}$ & $\rm{unw}$ & $2500\rho_{c}$\\
\end{tabular}
\end{center}
\raggedright{$^{a}$Cosmological parameters ($\Omega_{\rm{M}}$, $\Omega_{\rm{\Lambda}}$, $h$, $\rm{\sigma_8}$): \textit{WMAP}1 (0.25, 0.75, 0.73, 0.9); \textit{WMAP}3 (0.25, 0.75, 0.73, 0.9); \textit{WMAP}5 (0.258, 0.742, 0.72, 0.796); \textit{Planck} (0.307, 0.693, 0.678, 0.829); $\Lambda$CDM (0.3, 0.7, 0.7, 0.9).\\$^{b}$The halo shapes were determined at five different fraction of their virial radius in order to examine the radial dependence of $s$ with halo mass. We use this radial-shape relation to convert from 0.3$R_{\rm{vir}}$ to $R_{\rm{vir}}$.\\$^{c}$The analysis also includes 12 high-resolution haloes with more than 10$^6$ particles within the virial radius.}
\end{table*}


\section{Conclusions}
\label{sec:conclusions}
We have presented a detailed analysis on the triaxiality of dark matter haloes in four different computational volumes from the MultiDark \textit{Planck} simulation suite  \citep{Klypin2016}.  In particular, the Huge-MultiDark simulation allows us to characterize the shapes of cluster mass haloes with unprecedented statistics (almost 400000 haloes with masses above $2 \times 10^{14}~h^{-1}$M$_\odot$). Using the three simulations with smaller box volumes we extend our analysis to lower masses down to $3 \times 10^{11}~h^{-1}$M$_\odot$. The main results of our analysis can be  summarized  as follows.

\begin{itemize}

\item Dark matter haloes are well described by a triaxial model. In particular, due to interactions with other clusters or groups of galaxies, more massive haloes are less spherical than less massive ones (with a tendency of being prolate). Besides, massive haloes are also less spherical at high redshift essentially because they are affected by the direction of the last major merger and/or by the presence of filaments around them, and because massive haloes were formed later on.

\item Since the modelling of non-spherical objects is not simple and highly elongated haloes are mostly due to unrelaxed, we analyse separately relaxed and unrelaxed haloes based on the criteria proposed by \citet{Klypin2016}.

\item The iterative weighted and unweighted inertia tensors give similar results, while the inertia tensor in a spherical window is considerably higher than the formers.

\item Minor-to-major axis ratio can be expressed by a universal function in terms of the peak height $\nu$ (equation~\ref{eq:relation_axial}).  The peak height is related to the halo mass and incorporates the dependence on the cosmological parameters.

\item Rescaled minor-to-major axis ratio obtained from this functional form is well described by a lognormal distribution within the whole mass range considered.

\item Dark matter haloes are less spherical at inner radius, with axis ratios measured at $\rm{R_{500}}$ lower than those measured at $\rm{R_{vir}}$ on about 9 per cent.

\item The conditional intermediate-to-major axis ratio distribution ($p(q|s)$) is well fitted by a beta distribution, which depends only on the minor-to-major axis ratio (but not on the halo mass).

\item For $z=0$ haloes with masses larger than $10^{14}~h^{-1}$M$_\odot$ are dominated by prolate shapes (with fraction above 70 per cent), while only a few massive haloes are found to be triaxial or oblate in shape (with fraction below 20 and 5 per cent, respectively). At a higher redshift, the fraction of prolate massive haloes are even larger, with values of about 80 per cent.

\item The angle between the major axis measured at $\rm{R_{500}}$ and $\rm{R_{vir}}$ is below $30º$ for the 75 per cent of the triaxial and prolate haloes analysed, meaning that the direction of the major axis at $0.5 \rm{R_{vir}}$ aligns pretty well with the direction of the major axis at $\rm{R_{vir}}$.

\item Extremely low axis ratio are mostly due to massive and unrelaxed haloes; overall, the probability of finding a halo with $s < 0.1$ is $P(s<0.1) =10^{-3.20} \approx 6 \times 10^{-4}$,  When the analysis is restricted only to relaxed haloes these probabilities decrease down to $P(s<0.1) \approx 10^{-5.62}$ (about two haloes per million).

\end{itemize}

Finally, in Section \ref{sec:comparison} we  have compared our results with previous studies. Although there is an overall agreement with previous studies, we find a $\sim$10 per cent scatter mainly due to the different methods used to determine the axis ratios (i.e. inertia tensor, the radius at which the axis are measured) and the definition of the halo mass (spherical or elliptical). Additionally, in studies that include unrelaxed haloes, the fraction of unrelaxed haloes found in the simulated volumes can considerably affect the mass-shape scaling relation, leading to a lower normalization value of the axis ratios. Nevertheless, the best agreement is found for studies in which a large halo mass range is considered.

Using a suite of high resolution simulations the shape of dark matter haloes has been accurately calculated over five orders of magnitude in halo mass. These precise estimates are in general agreement with previous work done for smaller mass ranges with differences due to the different estimation of the shapes.

\section{Acknowledgements}
The authors gratefully acknowledge the Gauss Centre for Supercomputing e.V. (\url{www.gauss-centre.eu}) and the Partnership for Advanced Supercomputing in Europe (PRACE; \url{www.prace-ri.eu}) for funding the MultiDark simulation project by providing computing time on the GCS Supercomputer SuperMUC at the Leibniz Supercomputing Centre (LRZ; \url{www.lrz.de}) in Munich.  We would like to thank Peter Behroozi for his helpful comments on this paper and for making  the {\sc rockstar} halo finder software publicly  available.  We would like to acknowledge the anonymous referee for detailed comments that have contributed to improve the quality of this paper. The MultiDark data base was developed in cooperation with the Spanish MultiDark Consolider Project CSD2009-00064. The CosmoSim data base used in this paper is a service by the Leibniz-Institute for Astrophysics Potsdam (AIP). JV-F acknowledges financial support from MINECO (Spain) under the FPI grant Ref. BES-2010-039705 and travel grants Ref. EEBB-I-12-05072 and Ref. EEBB-I-14-08162. GY acknowledges financial support from MINECO (Spain) under research grants AYA2012-31101 and AYA2015-63810-P (MINECO/FEDER, EU). 

%
\bibliographystyle{mn2e}
\bibliography{shape} 

\begin{thebibliography}{}

\bibitem[\protect\citeauthoryear{{Allgood}, {Flores}, {Primack}, {Kravtsov},
  {Wechsler}, {Faltenbacher} \& {Bullock}}{{Allgood}
  et~al.}{2006}]{Allgood2006}
{Allgood} B.,  {Flores} R.~A.,  {Primack} J.~R.,  {Kravtsov} A.~V.,  {Wechsler}
  R.~H.,  {Faltenbacher} A.,    {Bullock} J.~S.,  2006, \mnras, 367, 1781

\bibitem[\protect\citeauthoryear{{Angulo}, {Springel}, {White}, {Jenkins},
  {Baugh} \& {Frenk}}{{Angulo} et~al.}{2012}]{Angulo2012}
{Angulo} R.~E.,  {Springel} V.,  {White} S.~D.~M.,  {Jenkins} A.,  {Baugh}
  C.~M.,    {Frenk} C.~S.,  2012, \mnras, 426, 2046

\bibitem[\protect\citeauthoryear{{Applegate}, {Mantz}, {Allen}, {der Linden},
  {Morris}, {Hilbert}, {Kelly}, {Burke}, {Ebeling}, {Rapetti} \&
  {Schmidt}}{{Applegate} et~al.}{2016}]{2016MNRAS.457.1522A}
{Applegate} D.~E.,  {Mantz} A.,  {Allen} S.~W.,  {der Linden} A.~v.,  {Morris}
  R.~G.,  {Hilbert} S.,  {Kelly} P.~L.,  {Burke} D.~L.,  {Ebeling} H.,
  {Rapetti} D.~A.,    {Schmidt} R.~W.,  2016, \mnras, 457, 1522

\bibitem[\protect\citeauthoryear{{Bailin} \& {Steinmetz}}{{Bailin} \&
  {Steinmetz}}{2005}]{Bailin2005}
{Bailin} J.,  {Steinmetz} M.,  2005, \apj, 627, 647

\bibitem[\protect\citeauthoryear{Bartelmann, Huss, Colberg, Jenkins \&
  Pearce}{Bartelmann et~al.}{1998}]{1998Bartelmann}
Bartelmann M.,  Huss A.,  Colberg J.~M.,  Jenkins A.,    Pearce F.~R.,  1998,
  Astron.Astrophys., 330, 1

\bibitem[\protect\citeauthoryear{{Bartelmann} \& {Schneider}}{{Bartelmann} \&
  {Schneider}}{2001}]{Bartelmann2001}
{Bartelmann} M.,  {Schneider} P.,  2001, \physrep, 340, 291

\bibitem[\protect\citeauthoryear{{Behroozi}, {Wechsler} \& {Wu}}{{Behroozi}
  et~al.}{2013}]{Behroozi2013}
{Behroozi} P.~S.,  {Wechsler} R.~H.,    {Wu} H.-Y.,  2013, \apj, 762, 109

\bibitem[\protect\citeauthoryear{{Bett}, {Eke}, {Frenk}, {Jenkins}, {Helly} \&
  {Navarro}}{{Bett} et~al.}{2007}]{Bett2007}
{Bett} P.,  {Eke} V.,  {Frenk} C.~S.,  {Jenkins} A.,  {Helly} J.,    {Navarro}
  J.,  2007, \mnras, 376, 215

\bibitem[\protect\citeauthoryear{{Binggeli}}{{Binggeli}}{1982}]{bingelli82}
{Binggeli} B.,  1982, \aap, 107, 338

\bibitem[\protect\citeauthoryear{{Bonamigo}, {Despali}, {Limousin}, {Angulo},
  {Giocoli} \& {Soucail}}{{Bonamigo} et~al.}{2015}]{Bonamigo2015}
{Bonamigo} M.,  {Despali} G.,  {Limousin} M.,  {Angulo} R.,  {Giocoli} C.,
  {Soucail} G.,  2015, \mnras, 449, 3171

\bibitem[\protect\citeauthoryear{{Bryan} \& {Norman}}{{Bryan} \&
  {Norman}}{1998}]{BN1998}
{Bryan} G.~L.,  {Norman} M.~L.,  1998, \apj, 495, 80

\bibitem[\protect\citeauthoryear{{Buote} \& {Canizares}}{{Buote} \&
  {Canizares}}{1992}]{buote92}
{Buote} D.~A.,  {Canizares} C.~R.,  1992, \apj, 400, 385

\bibitem[\protect\citeauthoryear{{Buote} \& {Canizares}}{{Buote} \&
  {Canizares}}{1996}]{buote96}
{Buote} D.~A.,  {Canizares} C.~R.,  1996, \apj, 457, 565

\bibitem[\protect\citeauthoryear{{Carroll}, {Press} \& {Turner}}{{Carroll}
  et~al.}{1992}]{Carroll1992}
{Carroll} S.~M.,  {Press} W.~H.,    {Turner} E.~L.,  1992, \araa, 30, 499

\bibitem[\protect\citeauthoryear{{Carter} \& {Metcalfe}}{{Carter} \&
  {Metcalfe}}{1980}]{carter80}
{Carter} D.,  {Metcalfe} N.,  1980, \mnras, 191, 325

\bibitem[\protect\citeauthoryear{{Cole} \& {Lacey}}{{Cole} \&
  {Lacey}}{1996}]{cole96}
{Cole} S.,  {Lacey} C.,  1996, \mnras, 281, 716

\bibitem[\protect\citeauthoryear{{Despali}, {Giocoli} \& {Tormen}}{{Despali}
  et~al.}{2014}]{Despali2014}
{Despali} G.,  {Giocoli} C.,    {Tormen} G.,  2014, \mnras, 443, 3208

\bibitem[\protect\citeauthoryear{{Despali}, {Tormen} \& {Sheth}}{{Despali}
  et~al.}{2013}]{Despali2013}
{Despali} G.,  {Tormen} G.,    {Sheth} R.~K.,  2013, \mnras, 431, 1143

\bibitem[\protect\citeauthoryear{{Dubinski} \& {Carlberg}}{{Dubinski} \&
  {Carlberg}}{1991}]{dubinski91}
{Dubinski} J.,  {Carlberg} R.~G.,  1991, \apj, 378, 496

\bibitem[\protect\citeauthoryear{{Evans} \& {Bridle}}{{Evans} \&
  {Bridle}}{2009}]{evans09}
{Evans} A.~K.~D.,  {Bridle} S.,  2009, \apj, 695, 1446

\bibitem[\protect\citeauthoryear{{Fabricant}, {Rybicki} \&
  {Gorenstein}}{{Fabricant} et~al.}{1984}]{fabricant84}
{Fabricant} D.,  {Rybicki} G.,    {Gorenstein} P.,  1984, \apj, 286, 186

\bibitem[\protect\citeauthoryear{{Faltenbacher}, {Allgood}, {Gottl{\"o}ber},
  {Yepes} \& {Hoffman}}{{Faltenbacher} et~al.}{2005}]{Faltenbacher2005}
{Faltenbacher} A.,  {Allgood} B.,  {Gottl{\"o}ber} S.,  {Yepes} G.,
  {Hoffman} Y.,  2005, \mnras, 362, 1099

\bibitem[\protect\citeauthoryear{{Ford}, {Van Waerbeke}, {Milkeraitis},
  {Laigle}, {Hildebrandt}, {Erben}, {Heymans} \& {Hoekstra}}{{Ford}
  et~al.}{2015}]{2015MNRAS.447.1304F}
{Ford} J.,  {Van Waerbeke} L.,  {Milkeraitis} M.,  {Laigle} C.,  {Hildebrandt}
  H.,  {Erben} T.,  {Heymans} C.,    {Hoekstra} H.,  2015, \mnras, 447, 1304

\bibitem[\protect\citeauthoryear{{Franx}, {Illingworth} \& {de Zeeuw}}{{Franx}
  et~al.}{1991}]{Franx1991}
{Franx} M.,  {Illingworth} G.,    {de Zeeuw} T.,  1991, \apj, 383, 112

\bibitem[\protect\citeauthoryear{{Frenk}, {White}, {Davis} \&
  {Efstathiou}}{{Frenk} et~al.}{1988}]{frenk88}
{Frenk} C.~S.,  {White} S.~D.~M.,  {Davis} M.,    {Efstathiou} G.,  1988, \apj,
  327, 507

\bibitem[\protect\citeauthoryear{{Gao}, {Navarro}, {Frenk}, {Jenkins},
  {Springel} \& {White}}{{Gao} et~al.}{2012}]{gao12}
{Gao} L.,  {Navarro} J.~F.,  {Frenk} C.~S.,  {Jenkins} A.,  {Springel} V.,
  {White} S.~D.~M.,  2012, \mnras, 425, 2169

\bibitem[\protect\citeauthoryear{{Giocoli}, {Meneghetti} \& et al.}{{Giocoli}
  et~al.}{2012a}]{Giocoli2012b}
{Giocoli} C.,  {Meneghetti} M.,    et al. 2012a, \mnras, 426, 1558

\bibitem[\protect\citeauthoryear{{Giocoli}, {Meneghetti} \& et al.}{{Giocoli}
  et~al.}{2012b}]{Giocoli2012a}
{Giocoli} C.,  {Meneghetti} M.,    et al. 2012b, \mnras, 421, 3343

\bibitem[\protect\citeauthoryear{{Groener} \& {Goldberg}}{{Groener} \&
  {Goldberg}}{2014}]{Groener2014}
{Groener} A.~M.,  {Goldberg} D.~M.,  2014, \apj, 795, 153

\bibitem[\protect\citeauthoryear{{Hoekstra}, {Herbonnet}, {Muzzin}, {Babul},
  {Mahdavi}, {Viola} \& {Cacciato}}{{Hoekstra}
  et~al.}{2015}]{2015MNRAS.449..685H}
{Hoekstra} H.,  {Herbonnet} R.,  {Muzzin} A.,  {Babul} A.,  {Mahdavi} A.,
  {Viola} M.,    {Cacciato} M.,  2015, \mnras, 449, 685

\bibitem[\protect\citeauthoryear{{Hopkins}, {Bahcall} \& {Bode}}{{Hopkins}
  et~al.}{2005}]{hopkins05}
{Hopkins} P.~F.,  {Bahcall} N.~A.,    {Bode} P.,  2005, \apj, 618, 1

\bibitem[\protect\citeauthoryear{{Jing} \& {Suto}}{{Jing} \&
  {Suto}}{2002}]{JS02}
{Jing} Y.~P.,  {Suto} Y.,  2002, \apj, 574, 538

\bibitem[\protect\citeauthoryear{{Kasun} \& {Evrard}}{{Kasun} \&
  {Evrard}}{2005}]{Kasun2005}
{Kasun} S.~F.,  {Evrard} A.~E.,  2005, \apj, 629, 781

\bibitem[\protect\citeauthoryear{{Kawahara}}{{Kawahara}}{2010}]{kawahara}
{Kawahara} H.,  2010, \apj, 719, 1926

\bibitem[\protect\citeauthoryear{{Kettula}, {Giodini}, {van Uitert},
  {Hoekstra}, {Finoguenov}, {Lerchster}, {Erben}, {Heymans}, {Hildebrandt},
  {Kitching}, {Mahdavi} \& {Mellier}}{{Kettula}
  et~al.}{2015}]{2015MNRAS.451.1460K}
{Kettula} K.,  {Giodini} S.,  {van Uitert} E.,  {Hoekstra} H.,  {Finoguenov}
  A.,  {Lerchster} M.,  {Erben} T.,  {Heymans} C.,  {Hildebrandt} H.,
  {Kitching} T.~D.,  {Mahdavi} A.,    {Mellier} Y.,  2015, \mnras, 451, 1460

\bibitem[\protect\citeauthoryear{{Klypin}, {Yepes}, {Gottl{\"o}ber}, {Prada} \&
  {He{\ss}}}{{Klypin} et~al.}{2016}]{Klypin2016}
{Klypin} A.,  {Yepes} G.,  {Gottl{\"o}ber} S.,  {Prada} F.,    {He{\ss}} S.,
  2016, \mnras, 457, 4340

\bibitem[\protect\citeauthoryear{{Knebe}, {Gill}, {Gibson}, {Lewis}, {Ibata} \&
  {Dopita}}{{Knebe} et~al.}{2004}]{Knebe2004}
{Knebe} A.,  {Gill} S.~P.~D.,  {Gibson} B.~K.,  {Lewis} G.~F.,  {Ibata} R.~A.,
    {Dopita} M.~A.,  2004, \apj, 603, 7

\bibitem[\protect\citeauthoryear{{Lau}, {Nagai} \& {Nelson}}{{Lau}
  et~al.}{2013}]{lau2013}
{Lau} E.~T.,  {Nagai} D.,    {Nelson} K.,  2013, \apj, 777, 151

\bibitem[\protect\citeauthoryear{{Ludlow}, {Bose}, {Angulo}, {Wang},
  {Hellwing}, {Navarro}, {Cole} \& {Frenk}}{{Ludlow} et~al.}{2016}]{Ludlow2016}
{Ludlow} A.~D.,  {Bose} S.,  {Angulo} R.~E.,  {Wang} L.,  {Hellwing} W.~A.,
  {Navarro} J.~F.,  {Cole} S.,    {Frenk} C.~S.,  2016, \mnras

\bibitem[\protect\citeauthoryear{{Medezinski}, {Umetsu}, {Okabe}, {Nonino},
  {Molnar}, {Massey}, {Dupke} \& {Merten}}{{Medezinski}
  et~al.}{2016}]{2016ApJ...817...24M}
{Medezinski} E.,  {Umetsu} K.,  {Okabe} N.,  {Nonino} M.,  {Molnar} S.,
  {Massey} R.,  {Dupke} R.,    {Merten} J.,  2016, \apj, 817, 24

\bibitem[\protect\citeauthoryear{{Meneghetti}, {Bolzonella}, {Bartelmann},
  {Moscardini} \& {Tormen}}{{Meneghetti} et~al.}{2000}]{2000Meneghetti}
{Meneghetti} M.,  {Bolzonella} M.,  {Bartelmann} M.,  {Moscardini} L.,
  {Tormen} G.,  2000, \mnras, 314, 338

\bibitem[\protect\citeauthoryear{{Meneghetti}, {Yoshida}, {Bartelmann},
  {Moscardini}, {Springel}, {Tormen} \& {White}}{{Meneghetti}
  et~al.}{2001}]{2001Meneghetti}
{Meneghetti} M.,  {Yoshida} N.,  {Bartelmann} M.,  {Moscardini} L.,  {Springel}
  V.,  {Tormen} G.,    {White} S.~D.~M.,  2001, \mnras, 325, 435

\bibitem[\protect\citeauthoryear{{Mo}, {Gonzalez}, {Jee}, {Massey}, {Rhodes},
  {Brodwin}, {Eisenhardt}, {Marrone}, {Stanford} \& {Zeimann}}{{Mo}
  et~al.}{2016}]{2016ApJ...818L..25M}
{Mo} W.,  {Gonzalez} A.,  {Jee} M.~J.,  {Massey} R.,  {Rhodes} J.,  {Brodwin}
  M.,  {Eisenhardt} P.,  {Marrone} D.~P.,  {Stanford} S.~A.,    {Zeimann}
  G.~R.,  2016, \apjl, 818, L25

\bibitem[\protect\citeauthoryear{{Mu{\~n}oz-Cuartas}, {Macci{\`o}},
  {Gottl{\"o}ber} \& {Dutton}}{{Mu{\~n}oz-Cuartas} et~al.}{2011}]{munozcuartas}
{Mu{\~n}oz-Cuartas} J.~C.,  {Macci{\`o}} A.~V.,  {Gottl{\"o}ber} S.,
  {Dutton} A.~A.,  2011, \mnras, 411, 584

\bibitem[\protect\citeauthoryear{{Murray}, {Power} \& {Robotham}}{{Murray}
  et~al.}{2013}]{Murray2013}
{Murray} S.~G.,  {Power} C.,    {Robotham} A.~S.~G.,  2013, Astronomy and
  Computing, 3, 23

\bibitem[\protect\citeauthoryear{{Natarajan} \& {Refregier}}{{Natarajan} \&
  {Refregier}}{2000}]{2000ApJ...538L.113N}
{Natarajan} P.,  {Refregier} A.,  2000, \apjl, 538, L113

\bibitem[\protect\citeauthoryear{Nelder \& Mead}{Nelder \&
  Mead}{1965}]{Nelder01011965}
Nelder J.~A.,  Mead R.,  1965, The Computer Journal, 7, 308

\bibitem[\protect\citeauthoryear{{Oguri}, {Bayliss}, {Dahle}, {Sharon},
  {Gladders}, {Natarajan}, {Hennawi} \& {Koester}}{{Oguri}
  et~al.}{2012}]{oguri12}
{Oguri} M.,  {Bayliss} M.~B.,  {Dahle} H.,  {Sharon} K.,  {Gladders} M.~D.,
  {Natarajan} P.,  {Hennawi} J.~F.,    {Koester} B.~P.,  2012, \mnras, 420,
  3213

\bibitem[\protect\citeauthoryear{{Oguri}, {Takada}, {Okabe} \& {Smith}}{{Oguri}
  et~al.}{2010}]{oguri10}
{Oguri} M.,  {Takada} M.,  {Okabe} N.,    {Smith} G.~P.,  2010, \mnras, 405,
  2215

\bibitem[\protect\citeauthoryear{{Oguri}, {Takada}, {Umetsu} \&
  {Broadhurst}}{{Oguri} et~al.}{2005}]{2005Oguri}
{Oguri} M.,  {Takada} M.,  {Umetsu} K.,    {Broadhurst} T.,  2005, \apj, 632,
  841

\bibitem[\protect\citeauthoryear{{Paz}, {Lambas}, {Padilla} \&
  {Merch{\'a}n}}{{Paz} et~al.}{2006}]{paz06}
{Paz} D.~J.,  {Lambas} D.~G.,  {Padilla} N.,    {Merch{\'a}n} M.,  2006,
  \mnras, 366, 1503

\bibitem[\protect\citeauthoryear{{Peebles}}{{Peebles}}{1969}]{Peebles1969}
{Peebles} P.~J.~E.,  1969, \apj, 155, 393

\bibitem[\protect\citeauthoryear{{Planck Collaboration}, {Ade}, {Aghanim},
  {Armitage-Caplan}, {Arnaud}, {Ashdown}, {Atrio-Barandela}, {Aumont},
  {Baccigalupi}, {Banday} \& et al.}{{Planck Collaboration}
  et~al.}{2014}]{planck_params}
{Planck Collaboration} {Ade} P.~A.~R.,  {Aghanim} N.,  {Armitage-Caplan} C.,
  {Arnaud} M.,  {Ashdown} M.,  {Atrio-Barandela} F.,  {Aumont} J.,
  {Baccigalupi} C.,  {Banday} A.~J.,    et al. 2014, \aap, 571, A16

\bibitem[\protect\citeauthoryear{{Press} \& {Schechter}}{{Press} \&
  {Schechter}}{1974}]{Press-Schechter1974}
{Press} W.~H.,  {Schechter} P.,  1974, \apj, 187, 425

\bibitem[\protect\citeauthoryear{{Redlich}, {Bartelmann}, {Waizmann} \&
  {Fedeli}}{{Redlich} et~al.}{2012}]{Redlich12}
{Redlich} M.,  {Bartelmann} M.,  {Waizmann} J.-C.,    {Fedeli} C.,  2012, \aap,
  547, A66

\bibitem[\protect\citeauthoryear{{Sayers}, {Golwala}, {Ameglio} \&
  {Pierpaoli}}{{Sayers} et~al.}{2011}]{sayers2011a}
{Sayers} J.,  {Golwala} S.~R.,  {Ameglio} S.,    {Pierpaoli} E.,  2011, \apj,
  728, 39

\bibitem[\protect\citeauthoryear{{Schneider}, {Frenk} \& {Cole}}{{Schneider}
  et~al.}{2012}]{Schneider2012}
{Schneider} M.~D.,  {Frenk} C.~S.,    {Cole} S.,  2012, \jcap, 5, 30

\bibitem[\protect\citeauthoryear{{Schrabback}, {Hilbert}, {Hoekstra}, {Simon},
  {van Uitert}, {Erben}, {Heymans}, {Hildebrandt} \& {Kitching}}{{Schrabback}
  et~al.}{2015}]{2015MNRAS.454.1432S}
{Schrabback} T.,  {Hilbert} S.,  {Hoekstra} H.,  {Simon} P.,  {van Uitert} E.,
  {Erben} T.,  {Heymans} C.,  {Hildebrandt} H.,    {Kitching} T.~D.,  2015,
  \mnras, 454, 1432

\bibitem[\protect\citeauthoryear{{Soucail}, {Fort}, {Mellier} \&
  {Picat}}{{Soucail} et~al.}{1987}]{soucail87}
{Soucail} G.,  {Fort} B.,  {Mellier} Y.,    {Picat} J.~P.,  1987, \aap, 172,
  L14

\bibitem[\protect\citeauthoryear{{Springel}}{{Springel}}{2005}]{Springel2005}
{Springel} V.,  2005, \mnras, 364, 1105

\bibitem[\protect\citeauthoryear{{Tinker}, {Kravtsov}, {Klypin}, {Abazajian},
  {Warren}, {Yepes}, {Gottl{\"o}ber} \& {Holz}}{{Tinker}
  et~al.}{2008}]{Tinker2008}
{Tinker} J.,  {Kravtsov} A.~V.,  {Klypin} A.,  {Abazajian} K.,  {Warren} M.,
  {Yepes} G.,  {Gottl{\"o}ber} S.,    {Holz} D.~E.,  2008, \apj, 688, 709

\bibitem[\protect\citeauthoryear{{Waizmann}, {Redlich} \&
  {Bartelmann}}{{Waizmann} et~al.}{2012}]{Waizmann12}
{Waizmann} J.-C.,  {Redlich} M.,    {Bartelmann} M.,  2012, \aap, 547, A67

\bibitem[\protect\citeauthoryear{{Waizmann}, {Redlich}, {Meneghetti} \&
  {Bartelmann}}{{Waizmann} et~al.}{2014}]{Waizmann14}
{Waizmann} J.-C.,  {Redlich} M.,  {Meneghetti} M.,    {Bartelmann} M.,  2014,
  \aap, 565, A28

\bibitem[\protect\citeauthoryear{{Warren}, {Quinn}, {Salmon} \&
  {Zurek}}{{Warren} et~al.}{1992}]{warren92}
{Warren} M.~S.,  {Quinn} P.~J.,  {Salmon} J.~K.,    {Zurek} W.~H.,  1992, \apj,
  399, 405

\bibitem[\protect\citeauthoryear{{Wojtak}}{{Wojtak}}{2013}]{Wojtak2013}
{Wojtak} R.,  2013, \aap, 559, A89

\bibitem[\protect\citeauthoryear{{Zentner}, {Berlind}, {Bullock}, {Kravtsov} \&
  {Wechsler}}{{Zentner} et~al.}{2005}]{Zentner2005}
{Zentner} A.~R.,  {Berlind} A.~A.,  {Bullock} J.~S.,  {Kravtsov} A.~V.,
  {Wechsler} R.~H.,  2005, \apj, 624, 505

\end{thebibliography}
\appendix 
\section{Density peak height}\label{sec:nu}
The density peak height $\nu$ for a virialized halo with mass $M$ at redshift $z$ for a
given cosmological model is given by:

\begin{equation}
\nu \equiv \frac{\delta_c(z)}{\sigma(M)},
\end{equation}
where $\delta_c(z)$  is the  critical  overdensity  of the  spherical collapse  model (i.e. the  initial density required  for  a
fluctuation  to  collapse  at  redshift  $z$) and $\sigma(M)$ is the variance in the initial density field smoothed on a linear scale  $R$ (which corresponds to the radius of a  uniform sphere of mass  $M$).

The critical  overdensity can  be rewritten as the collapse overdensity at present time ($\delta_c(z=0) = \delta_c \approx 1.69$) rescaled to a given time as follows: $\delta_c(z) =  \delta_c / D(z)$, with $D(z)$ being the linear growth rate of a density fluctuation normalized to unity at the present time. 

For the computation of $D(z)$ we use the approximation given by \citet{Carroll1992}, which can be expressed as $D(z) \propto g(z) / (1+z)$ with

\begin{equation}
	g(z) = \frac{5/2\, \Omega_{\rm{m}}(z)}{\Omega_{\rm{m}}^{4/7} - \Omega_\Lambda(z) + \left[1+\Omega_{\rm{m}}(z)/2 \right] \left[ 1+\Omega_\Lambda(z)/70 \right] }.
	\label{eq:growingmode}
\end{equation}

The variance in the initial density field, $\sigma(M)$, is defined from the power spectrum as:
\begin{equation}
	\sigma^2(M) = \frac{1}{2\pi^2} \int_0^\infty P(k) \tilde{W}^2(kR) k^2 \text{d} k;
	\label{eq:sigma}
\end{equation}
where $\tilde{W}$ is the Fourier transform of the top-hat window function. For a given cosmology, both the power spectrum ($P(k)$) and the variance in the initial density field ($\sigma(M)$) can be computed using a software like
\textsc{HMFcalc} \citep{Murray2013}.

On one hand, the overdensity $\delta_c$, which depends only on redshift and not on the mass, increases with redshift due to $D(z)$; while, on the other  hand, the $\sigma(M)$ depends on the mass but not on redshift (since $\sigma(M) = \sigma(M,z=0)$ by definition). Therefore, to compute the peak height $\nu$ for a halo of  mass $M$ only the linear growth rate $D(z)$ and the initial power spectrum $P(k)$ are needed.

\begin{figure}
\includegraphics[width=8.5cm,angle=0.0]{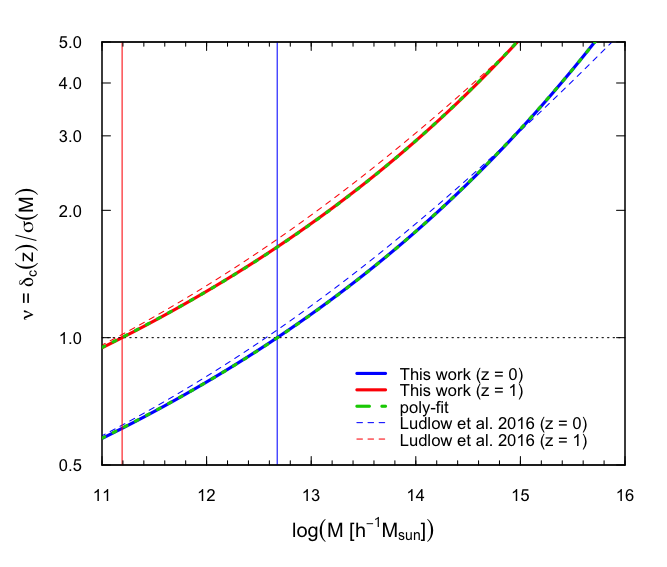}
\caption{Peak height as a function of the virial halo mass for the \textit{Planck} cosmology. Solid lines correspond with the results obtained in this work, while dashed lines show the findings of \citet{Ludlow2016}. Blue and red lines indicate the corresponding redshift $z=0$ and 1, respectively. The dashed green lines correspond to the best fit with equation~(\ref{eq:numass}). The vertical lines indicate the characteristic masses at each redshift given by $\nu (z, M_{*}) = 1$.}
\label{fig:numass}
\end{figure}

Additionally, we proposed a relation between the halo mass $M$ and the peak height $\nu$, for the cosmological parameters given in Section~\ref{sec:catalogs}. At present time, the relation between $\nu (z=0) = \nu_0$ and the halo mass (in log--log space) can be expressed as a third order polynomial:
	
\begin{equation}
\mathrm{log} ~ \nu_0 = n_0 + n_1~x + n_2~x^2 + n_3~x^3\;,
\label{eq:numass}
\end{equation}
with $x = \rm{log}~M$. As can be clearly seen in Fig.~\ref{fig:numass}, this relation gives an excellent fit within the whole mass range  studied  here with best-fitting parameters: $n_0 = -2.63 \pm 0.04$, $n_1 = 0.504 \pm 0.009$, $n_2 = -0.0434 \pm 0.0007$ and $n_3 = 0.00158 \pm 0.00002$. This result is in good agreement with the functional form proposed by \citet{Ludlow2016} for a similar set of cosmological parameters.

\label{lastpage}
\end{document}